\documentclass[preprint,aps,12pt]{revtex4}

\usepackage{graphicx}
\usepackage{dcolumn}
\usepackage{bm}
\usepackage{color}
\usepackage{amsmath}
\usepackage{amssymb}

\def\slashchar#1{\setbox0=\hbox{$#1$}           
   \dimen0=\wd0                                 
   \setbox1=\hbox{/} \dimen1=\wd1               
   \ifdim\dimen0>\dimen1                        
      \rlap{\hbox to \dimen0{\hfil/\hfil}}      
      #1                                        
   \else                                        
      \rlap{\hbox to \dimen1{\hfil$#1$\hfil}}   
      /                                         
   \fi}

\begin{document}

\preprint{TUW-05-17}

\title{Nonperturbative renormalization of $\Phi$-derivable approximations in theories with fermions}

\author{Urko Reinosa}
\email{u.reinosa@thphys.uni-heidelberg.de}
  \affiliation{Institut f$\ddot{u}$r Theoretische Physik, Technische Universit$\ddot{a}$t Wien\\
Wiedner Hauptstrasse 8-10/136, A-1040 Wien, Austria.}

\date{\today} 

\begin{abstract}
We show how to renormalize $\Phi$-derivable approximations in a theory with a fermionic field coupled to a self-interacting scalar field through a Yukawa interaction. The nonperturbative renormalization concerns the self-interaction coupling of the scalar field which is renormalized via a set of nested Bethe-Salpeter equations for the scalar and fermionic four-point functions. We use this information to construct explicit finite equations of motion in the symmetric phase. We work in the context of equilibrium quantum field theory and show that the renormalization can be carried out without introducing temperature dependent counterterms. 
\end{abstract}

\pacs{Valid PACS appear here}
\maketitle

\section{Introduction}

Formulation of quantum field theory in terms of the two-particle-irreducible (2PI) effective action has recently regained interest in the context of QCD. It leads for example to a good description of the entropy of the quark-gluon plasma from very high temperatures down to temperatures of the order of three times the critical temperature \cite{Blaizot:1999ip,Blaizot:1999ap,Blaizot:2000fc}. The 2PI effective action was introduced in the context of the non-relativistic many body problem \cite{Luttinger:1960ua,Baym:1962sx,DeDom:1964vz} and was then extended to quantum field theory in \cite{Cornwall:1974vz}. It aims at parametrising physical quantities in terms of one- and two-point functions. For quantities dominated by fluctuations encoded in one- and two-point functions, an expansion of the 2PI effective action is then expected to show better convergence properties than a perturbative (1PI) expansion. This can be explicitly checked in equilibrium within scalar theories \cite{Berges:2004hn} and there are evidences \cite{Blaizot:1999ip,Blaizot:1999ap,Blaizot:2000fc,Blaizot:2005wr} that the same phenomenon could cure the poor convergence of perturbative expansions \cite{Arnold:1995eb,Zhai:1995ac,Braaten:1996jr,Kajantie:2002wa} for particular thermodynamic quantities in QCD. Many other approaches aim at improving the convergence of perturbation theory using propagator resummation. This is for example the case of the so-called {\it screened perturbation theory} for scalar fields \cite{Karsch:1997gj,Andersen:2000yj} or its generalization to QCD \cite{Andersen:1999fw,Andersen:2002ey,Andersen:2003zk}. These other approaches are based on a resummation of the leading contribution to the self-energy for the soft modes (Hard Thermal Loops \cite{Braaten:1992gm,Frenkel:1992ts}). However they also include these corrections at hard scales leading to spurious contributions, in particular artificial temperature dependent ultraviolet divergences. In contrast, the 2PI effective action treats soft and hard scales on a different footing thus preventing the presence of unwanted temperature dependent singularities. Finally the use of the 2PI effective action goes now beyond the description of systems in equilibrium since it has been successfully applied to describe late time dynamics of quantum fields out-of-equilibrium \cite{Berges:2001fi,Berges:2000ur,Aarts:2002dj,Berges:2002wr,Berges:2002cz,Berges:2004ce,Cooper:2002qd,Juchem:2003bi,Arrizabalaga:2004iw,Arrizabalaga:2005tf} and may then be useful in understanding equilibration properties of the system created in ultra-relativistic heavy-ion collisions.

In order to do calculations within the 2PI framework, one has to devise approximations of the 2PI effective action. A particular class of approximations is obtained by considering diagrammatic truncations of the skeleton expansion \cite{Cornwall:1974vz} for the 2PI effective action. These approximations, known as $\Phi$-derivable \cite{Baym:1962sx}, are important since they respect global symmetries \cite{Luttinger:1960ua,Knoll:2001jx,vanHees:2002bv} and thus insure the presence of conserved quantities (for instance, energy conservation is crucial in order to properly simulate the time evolution of quantum fields). There are however two major drawbacks when constructing $\Phi$-derivable approximations. The first concerns any renormalizable theory and has to do with the possibility or not to remove UV divergences within a given $\Phi$-derivable approximation. The second concerns gauge theories and how to preserve gauge invariance. The difficulties are related to the fact that a given truncation of the effective action only resums particular topologies and it is not clear how fundamental properties such as finiteness or gauge invariance manifest themselves within such a subclass of diagrams. Recent progresses have been made concerning renormalization in the case of scalar theories \cite{vanHees:2001ik,vanHees:2001pf,Reinosa:2003qa,Blaizot:2003br,Blaizot:2003an,Reinosa:2004bn,Berges:2005hc,Cooper:2004rs,Cooper:2005vw}. For such theories, divergent topologies can be self-consistently renormalized by means of zero temperature counterterms. The question remains however totally opened to know what can be said about renormalization in the case of gauge theories. Two new features need to be understood: How to renormalize $\Phi$-derivable approximations coupling different type of fields and the interplay between renormalization and symmetry. We choose here to address the simpler one: As a first step we consider a situation with the same power counting rules as QED but without the difficulties related to the gauge symmetry. The aim is to see if the divergent topologies can be again self-consistently renormalized in a situation with two kind of fields. We thus consider the case of a fermionic field coupled to a self-interacting scalar field via a Yukawa interaction. We work in the context of equilibrium quantum field theory and discuss renormalization of the equations of motion in the symmetric phase. The broken phase will be dealt with in a subsequent work.

Section II introduces the 2PI effective action together with the relevant equations involved in the renormalization process. The result is indeed an extension of what is done in \cite{vanHees:2001ik,vanHees:2001pf,Blaizot:2003br,Blaizot:2003an}. In particular, renormalization of singularities involving four scalar fields relies on the same Bethe-Salpeter equation. The new feature is that, the kernel of the latter arises itself from a second Bethe-Salpeter equation for a four-point function involving only fermionic fields. This set of nested Bethe-Salpeter equations is the key to disentangle the coupling between scalar and fermionic degrees of freedom in the analysis of UV divergences. In section III, we use this information to renormalize the equations of motion in the one-loop approximation. Finally, Section IV is devoted to the generalization to any loop order. We provide a systematic rule for renormalization. We conclude by saying a few words on the difficulties met when extending these results to Abelian gauge theories.

\section{Divergent topologies of the 2PI effective action}

We consider a fermionic field $\psi$ coupled to a self-interacting scalar field $\varphi$ ($\varphi^4$ theory) via a Yukawa interaction, in four dimensions. We consider for the moment the unrenormalized ({\it bare}) action which reads
\begin{equation}
S=\int d^4x\,\left\{\bar\psi(i\slash\hspace{-2.1mm}\partial-m)\psi-g\bar{\psi}\psi\varphi+\frac{1}{2}(\partial\varphi)^2-\frac{1}{2}M^2\varphi^2-\frac{1}{4!}\lambda\varphi^4\right\}\,.
\end{equation} 
The time integral runs from $0$ to $-i\beta$ where $\beta=1/T$ denotes the inverse temperature. This theory is perturbatively renormalizable: The couplings in the action are dimensionless or, equivalently, the superficial degree of divergence $\delta$ of any Feynman diagram does not depend on the number of vertices but only on the number of scalar ($E_\varphi$) and fermionic ($E_\psi$) external legs: $\delta=4-E_\varphi-(3/2)E_\psi$. The theory has to be regularized so that one can write meaningful expressions in auxiliary steps before renormalization. In this paper, we use dimensional regularization which amounts to working in dimension $d=4-2\epsilon$. In that case, the couplings acquire a dimension which we parametrize by a renormalization mass scale $\mu$: $g\rightarrow\mu^{\epsilon}g$ and $\lambda\rightarrow\mu^{2\epsilon}\lambda$.

We use the Minkowski metric $(+,-,-,-)$ and work in Fourier space. We denote 4-momenta by capital letters $Q=(q_0,q)$. At finite temperature in imaginary time, the frequency integrals or sums are performed over the imaginary axis $Q=(q_0=i\omega,q)$.  We use the following notations:
\begin{eqnarray}
\int_{Q}^{\rm (T)}f(Q) & = & \frac{1}{\beta}\sum_n\int\frac{d^{d-1}q}{(2\pi)^{d-1}}\, f(i\omega_n,q)\,,\nonumber\\
\int_Q f(Q) & = & \int_{-\infty}^{\infty}\frac{d\omega}{2\pi}\int\frac{d^{d-1}q}{(2\pi)^{d-1}}\, f(i\omega,q)\,,\nonumber\\
\int_{\tilde Q} f(\tilde Q) & = & \int_{-\infty}^{\infty}\frac{dq_0}{2\pi}\int\frac{d^{d-1}q}{(2\pi)^{d-1}}\, f(q_0,q)\,.
\end{eqnarray}

\subsection{2PI effective action and equations of motion}
\begin{figure}[htbp]
\begin{center}
\includegraphics[width=12cm]{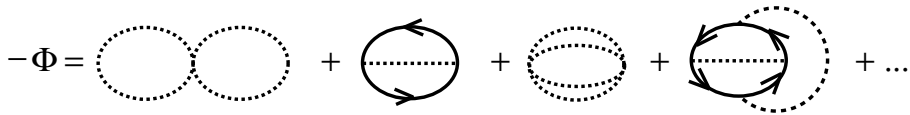}
\caption{First contributions to $\Phi[S,D;g,\lambda]$. Dashed and solid lines denote scalar and fermionic propagators respectively. The vertices are bare, tree level vertices.\label{fig:Phi}}
\end{center}
\end{figure}

In the symmetric phase, the mean values of both fields vanish and in order to study thermodynamic properties of the system, it is enough to consider the 2PI effective action as a functional of the full propagators $S$ and $D$. It is given by the expression
\begin{equation}\label{eq:2PIEA}
\Gamma_{\mbox{\tiny 2PI}}[S,D]=-\mbox{Tr}\,\ln S^{-1}+\mbox{Tr}\,\Sigma S+\frac{1}{2}\mbox{Tr}\,\ln D^{-1}-\frac{1}{2}\mbox{Tr}\,\Pi D+\Phi[S,D;g,\lambda]\,.
\end{equation}
The trace symbol {\it Tr} includes $(d-1)$-momentum integrals, Matsubara sums, and sums over spin indices. The self-energies $\Sigma$ and $\Pi$ are related to the propagators by the usual Schwinger-Dyson equations:
\begin{equation}
S_{\alpha\beta}^{-1}(P)=-\slash\hspace{-2.5mm}P+m+\Sigma_{\alpha\beta}(P)\,,\quad D^{-1}(K)=K^2+M^2+\Pi(K)\,.
\end{equation}
The functional $\Phi[S,D;g,\lambda]$ is given (up to a minus sign) by the sum of zero-leg two-particle-irreducible diagrams drawn with propagators $S$ and $D$, and vertices $g$ and $\lambda$. In Fig.~\ref{fig:Phi}, we represent the first diagrams contributing to $\Phi$. The functionals $\Phi$ and $\Gamma_{\mbox{\tiny 2PI}}$ depend a priori on the regulator $\epsilon$ and on the renormalization scale $\mu$ even if we do not make it explicit in our notation.

The physical propagators are given by the stationary condition on $\Gamma_{\mbox{\tiny 2PI}}[S,D]$ which is suitably translated into a set of equations of motion:
\begin{equation}\label{eq:gap}
\Sigma_{\alpha\beta}=-\frac{\delta \Phi}{\delta S_{\beta\alpha}}\,,\quad \Pi=2\frac{\delta \Phi}{\delta D}\,.
\end{equation}
These equations state that, in order to obtain the stationary self-energies, one has simply to remove respectively a fermionic or a scalar line from the diagrams in $\Phi$. By evaluating the 2PI effective action (\ref{eq:2PIEA}) at its stationary point, one obtains an expression for the pressure of the system in terms of the full propagators $S$ and $D$. Other thermodynamic quantities can be obtained by taking derivatives of the pressure. 

\begin{figure}[htbp]
\begin{center}
\includegraphics[width=10cm]{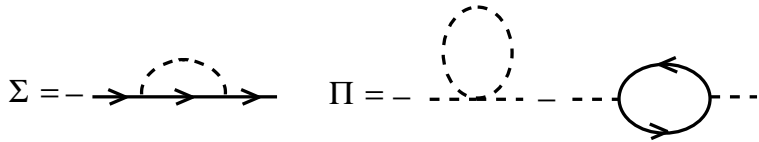}
\caption{Equations of motion at one-loop level (the lines represent full propagators).\label{fig:gap}}
\end{center}
\end{figure}

$\Phi$-derivable approximations are defined by selecting a class of diagrams contributing to $\Phi$ in Eq.~(\ref{eq:2PIEA}) and changing the equations of motion (\ref{eq:gap}) accordingly. If the diagramatic truncation of $\Phi$ is done according to the number of loops we refer to it as the 2PI-loop expansion. Further approximations can be build on top of the 2PI-loop expansion such as the ones used in \cite{Blaizot:1999ip,Blaizot:1999ap,Blaizot:2000fc} or in \cite{Braaten:2001vr,Braaten:2001en,Andersen:2004re} (see also \cite{Roder:2005qy,Roder:2005vt}). They have the advantage to be easier to implement numerically than a full self-consistent calculation. However the first approach is not systematic and the second is based on an expansion which leads to unbalanced temperature dependent singularities. In contrast the 2PI-loop expansion is systematic and as we will show can be renormalized without introducing artificial counterterms. Before dealing with general 2PI-loop approximations in section \ref{sec:higher_loops}, we illustrate the renormalization procedure at one-loop level \footnote{The number of loops refers to the equation of motion.} for which the equations of motion read
\begin{eqnarray}\label{eq:gap1loop}
\Sigma_{\alpha\beta}(P) & = & -g^2\int^{(T)}_Q\,D(Q)S_{\alpha\beta}(Q+P)\,,\nonumber\\
\Pi(K) & = & \frac{1}{2}\lambda\int^{(T)}_Q\,D(Q)+g^2\int^{(T)}_Q\,\mbox{tr}\Big\{S(Q)S(Q+K)\Big\}\,.
\end{eqnarray}
The notation $\int^{(T)}_Q$ includes finite temperature contributions via  Matsubara sums. The trace {\it tr} is taken over spin indices and we have taken into account the minus sign arising from the fermion loop. These equations are represented diagrammatically in Fig. \ref{fig:gap}.

\subsection{Iterations and divergences}\label{sec:iterations}
The aim of this section is to reveal some of the topologies which play a role in the process of renormalization at one-loop level. This simple example is indeed very fruitful since the structure of the equations that it leads to plays a general role (see section \ref{sec:higher_loops}). Most of the ideas are those already described in \cite{Blaizot:2003an}. The strategy to study divergences in the equations of motion is to iterate them, to express formally their solution in terms of an infinite sum of perturbative diagrams and to ask whether it is possible to tune the counterterms in order to absorb the singularities \footnote{This is indeed a necessary condition for the renormalization program to work at all.}. In order to do this consistently, one should proceed order by order in $\lambda$ and $g$. Here we just give examples of the divergent topologies that one reveals by this procedure and claim that there are no other ones. Later we prove this statement by providing renormalized equations of motion.

\begin{figure}[htbp]
\begin{center}
\includegraphics[width=12cm]{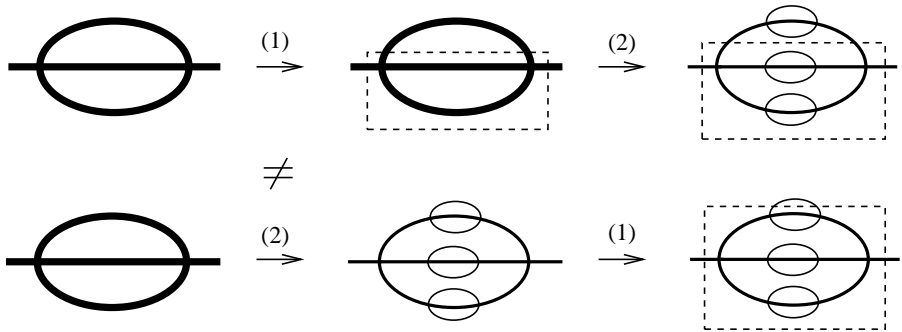}
\caption{The BPH procedure on resummed diagrams misses singularities. This is simply because the two operations {\it drawing boxes} (1) and {\it iterating diagrams} (2) do not commute as illustrated in the figure. The first line shows the coupling singularities which are accounted for by a BPH analysis regardless of the content of the propagator. The second line shows what happens if we first look inside the propagator and then apply the BPH analysis. There are clearly more divergent topologies in the second approach.\label{fig:exemple}}
\end{center}
\end{figure}

At one-loop level we will be only concerned by singularities involving four scalar fields (contributing to the scalar coupling counterterm: $\delta\lambda$). Field strength singularities are dealt with when treating the asymptotic equations of motion (see section \ref{sec:asymptotic}). Mass counterterms do not play a role since we restrict ourselves to the massless theory in dimensional regularization \footnote{A nonzero mass can be easily implemented following \cite{Blaizot:2003an}.}. The Yukawa coupling counterterm $\delta g$ is treated in section \ref{sec:higher_loops} when dealing with higher loops where its renormalization becomes important. At the level of the one-loop approximation, $\delta g=0$. Things are different for the scalar coupling counterterm $\delta\lambda$ which receives contributions to all orders. As was shown in \cite{Blaizot:2003an} in the case of the pure $\varphi^4$ scalar theory, an analysis of divergences based on BPH (Bogoliubov-Parasiuk-Hepp) boxes \cite{Bogol,Hepp:1966eg,Collins} with full propagators misses a large amount of coupling singularities. The idea is that if one views the propagator as build from a resummation of diagrams, one creates new boxes which are not taken into account by the BPH procedure applied to resummed diagrams. In Fig.~\ref{fig:exemple}, we give an illustration of such a phenomenon in the scalar $\varphi^4$ theory. We will show that these singularities can be accounted for by a (temperature independent) redefinition of the scalar coupling counterterm in front of the {\it tadpole} diagram (see for instance Fig. \ref{fig:gap}). As we will show, this redefinition is governed by a Bethe-Salpeter equation \footnote{An alternative would be to use a temperature dependent renormalization scheme and to absorb the singularities in a temperature dependent mass counterterm. One would then have to analyze the running of the renormalized parameters with temperature by following the kind of analysis suggested in \cite{Jakovac:2004ua}. The related renormalization group equations should involve the same Bethe-Salpeter equation as the one entering $\delta\lambda$ in the temperature independent renormalization scheme.}.

\begin{figure}[htbp]
\begin{center}
\includegraphics[width=10cm]{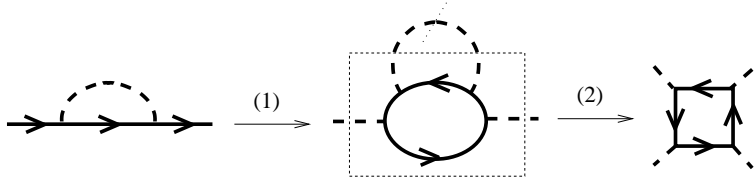}
\caption{UV singularity generated by plugging the fermion self-energy to lower order into the equation of motion for the scalar propagator. The lines represent perturbative propagators.\label{fig:singularity1}}
\end{center}
\end{figure}

\begin{figure}[htbp]
\begin{center}
\includegraphics[width=14cm]{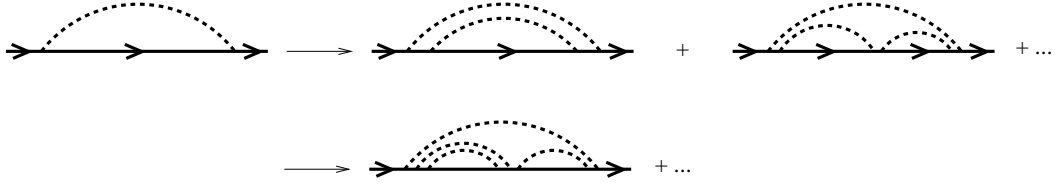}
\caption{Rainbow diagrams generated by iterating the fermion equation of motion with fixed scalar propagator ($S=S_0$).\label{fig:rainbow}}
\end{center}
\end{figure}

\begin{figure}[htbp]
\begin{center}
\includegraphics[width=16cm]{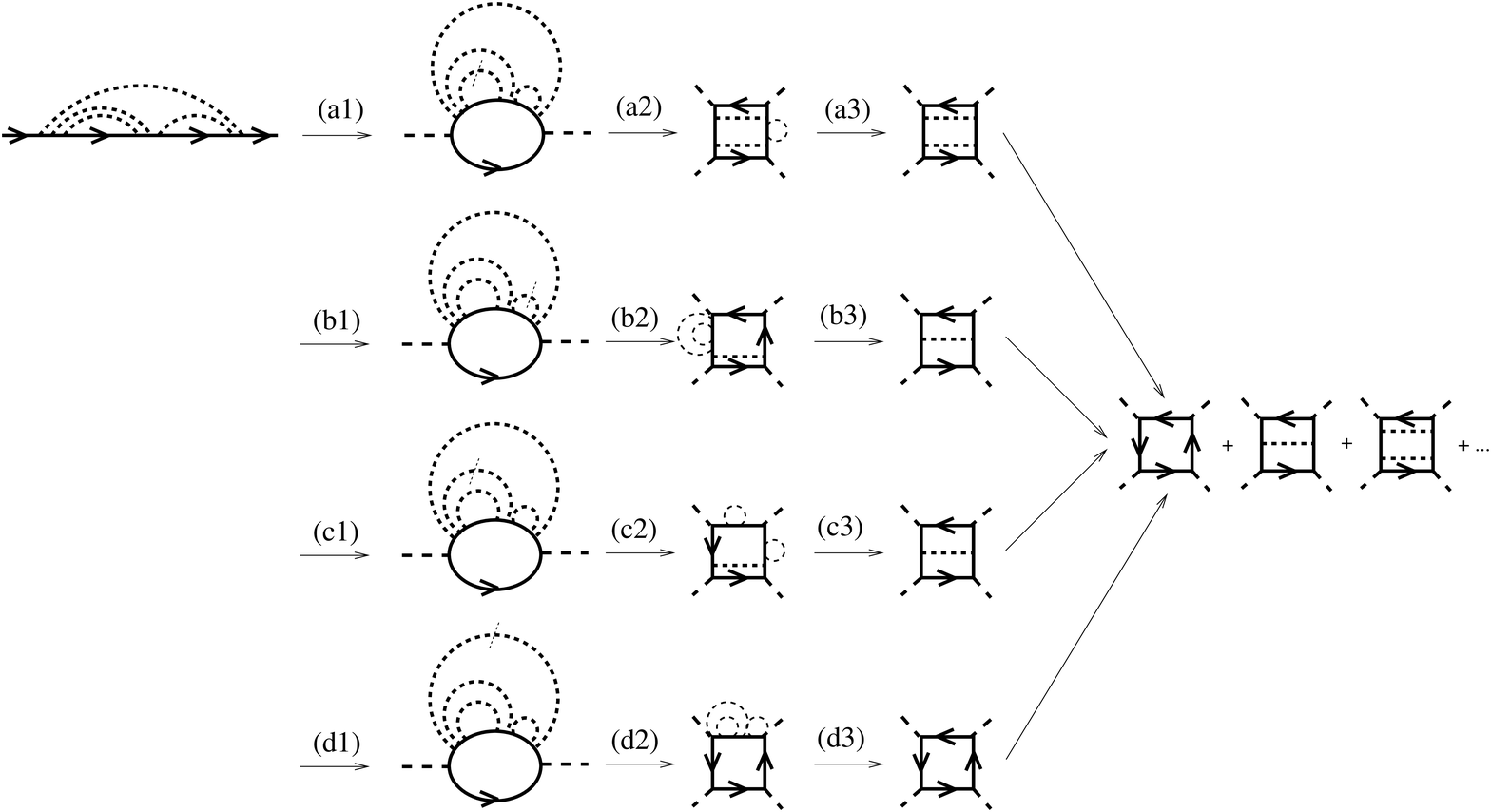}
\caption{The rainbow diagrams modify the singularity of Fig.~\ref{fig:singularity1}. All fermionic lines represent perturbative propagators $S_0$. Only in the last step (3) these are resummed into $S$.\label{fig:singularity2}}
\end{center}
\end{figure}

In our present example, the singularities appear in the equation of motion for the scalar propagator after iterating the coupled equations of motion a certain number of times. By iterations we mean that, knowing the solutions of these equations $\Sigma^{(n)}$ and $\Pi^{(n)}$ to a given accuracy, we can generate new contributions by plugging the related propagators $S^{-1}_{(n)}=S_0^{-1}+\Sigma^{(n)}$ and $G_{(n)}^{-1}=G_0^{-1}+\Pi^{(n)}$ into the equations of motion and expanding perturbatively according to
\begin{eqnarray}
S_{(n)} & = & S_0-S_0\Sigma^{(n)}S_0+S_0\Sigma^{(n)}S_0\Sigma^{(n)}S_0+\dots\nonumber\\
G_{(n)} & = & G_0-G_0\Pi^{(n)}G_0+G_0\Pi^{(n)}G_0\Pi^{(n)}G_0+\dots
\end{eqnarray}
 The simplest such a divergence appears by considering the 0th iteration of the equation of motion for the fermionic propagator (i.e. the perturbative one-loop contribution) and plugging it into the equation of motion for the scalar propagator. After opening the scalar line (opening a line corresponds, in the BPH language, to drawing a box which pulls this line apart) one ends up with the diagram in the right hand side of Fig.~\ref{fig:singularity1}, which is logarithmically divergent according to its superficial degree of divergence ($\delta=0$).
A simple generalization of this construction consists in first iterating the equation of motion for the fermionic propagator by keeping the scalar propagator fixed to its free value $S_0$. We generate the rainbow diagrams of Fig.~\ref{fig:rainbow} which, once plugged into the equation of motion for the scalar propagator, lead to the singularities of Fig.~\ref{fig:singularity2}, which generalize the singularity in Fig.~\ref{fig:singularity1}. The first two steps are the same than in Fig.~\ref{fig:singularity1}: After iterating (step 1), one selects a scalar line to isolate a singularity with four scalar legs (step2). Step 3 only resums the free fermionic propagators $S_0$ into the full propagator $S$. The final result of this procedure is the general structure on the right of Fig.~\ref{fig:singularity2} which generalizes the one in Fig.~\ref{fig:singularity1} by summing all the possible ladders made of scalars rungs.

\begin{figure}[htbp]
\begin{center}
\includegraphics[width=12cm]{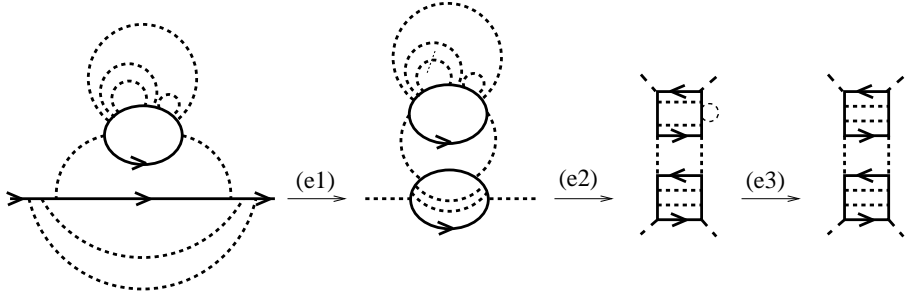}
\caption{Structures such as the ones depicted in Fig.~\ref{fig:singularity2} are concatenated through the iteration of the equations of motion.\label{fig:singularity3}}
\end{center}
\end{figure}

\begin{figure}[htbp]
\begin{center}
\includegraphics[width=3cm]{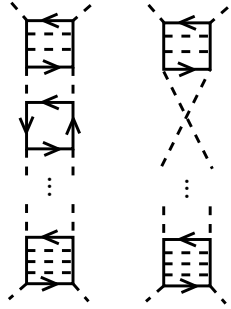}
\caption{General structure of scalar coupling singularities in the one-loop approximation.\label{fig:singularities}}
\end{center}
\end{figure}

One can finally repeat the same construction by considering corrections to the scalar propagator. If we consider for example the scalar self-energy correction that we discussed in the previous paragraph and take it into account in the generation of rainbows, one obtains the divergences depicted in Fig.~\ref{fig:singularity3} which are nothing but a concatenation of those depicted in Fig.~\ref{fig:singularity2}. The general divergent topology contributing to $\delta\lambda$ is then the one depicted in the first diagram of Fig.~\ref{fig:singularities}. The {\it tadpole} diagram brings trivial modifications to this topology (see the second diagram of Fig.~\ref{fig:singularities}). We claim here that these are the only singularities generated by the equations of motion which contribute to $\delta\lambda$. We give an algebraic proof of this statement when renormalizing the equations of motion in section \ref{sec:renormalization}.

\subsection{Kernels and Bethe-Salpeter equations}
\begin{figure}[htbp]
\begin{center}
\includegraphics[width=12cm]{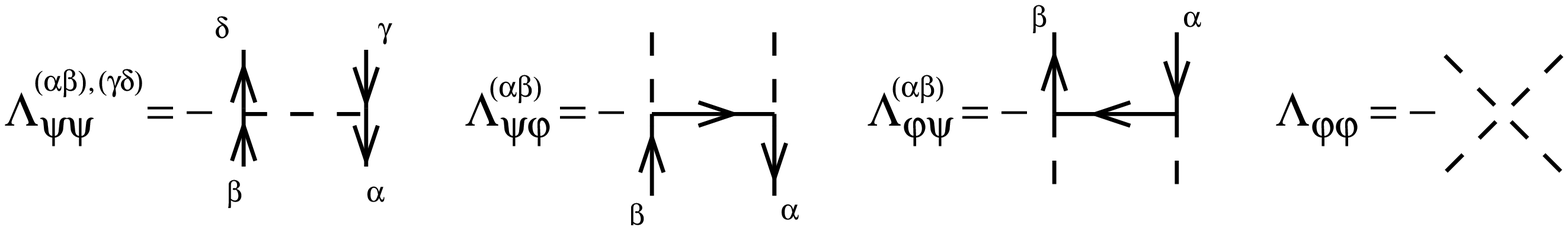}
\caption{Basic kernels in the one-loop approximation.\label{fig:kernels}}
\end{center}
\end{figure}

\begin{figure}[htbp]
\begin{center}
\includegraphics[width=14cm]{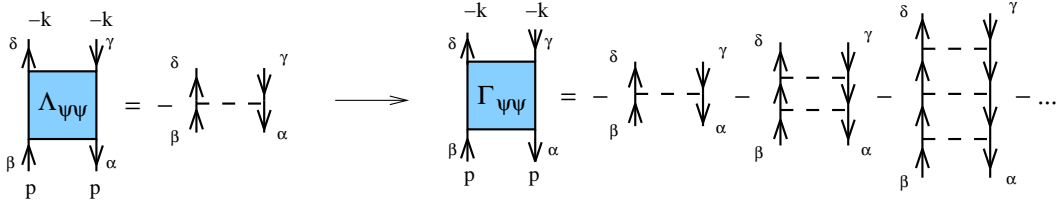}
\caption{First Bethe-Salpeter equation.\label{fig:Lambda_tilde}}
\end{center}
\end{figure}

\begin{figure}[htbp]
\begin{center}
\includegraphics[width=10cm]{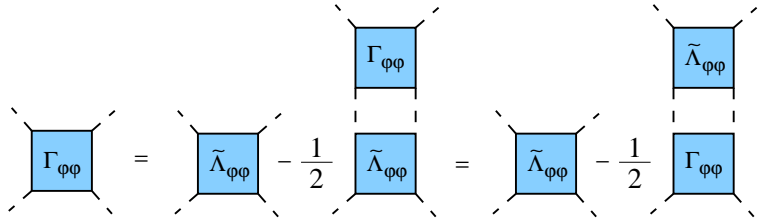}
\caption{Second Bethe-Salpeter equation. It encodes the coupling singularities of the equation of motion for the scalar propagator.\label{fig:BS}}
\end{center}
\end{figure}

Let us for the moment analyze in further details the topologies described in the previous section. We show that it is possible to simply encode them using nested Bethe-Salpeter equations. To this purpose, it is convenient to introduce the following 2PI {\it kernels} (the particular ordering of labels is introduced for later convenience):
\begin{equation}\label{eq:kernels}
\Lambda^{(\alpha\beta),(\gamma\delta)}_{\psi\psi}=-\frac{\delta^2\Phi}{\delta S_{\beta\alpha}\delta S_{\gamma\delta}}\,,\quad\Lambda^{(\alpha\beta)}_{\psi\varphi}=-2\frac{\delta^2\Phi}{\delta S_{\beta\alpha}\delta D}\,,\quad\Lambda^{(\alpha\beta)}_{\varphi\psi}=-2\frac{\delta^2\Phi}{\delta D\delta S_{\alpha\beta}}\,,\quad\Lambda_{\varphi\varphi}=4\frac{\delta^2\Phi}{\delta D\delta D}\,.
\end{equation}
The numerical factors are introduced so that the kernels represent (up to a minus sign) the sum of related diagrams. In Fig. \ref{fig:kernels}, we represent them at one-loop level.
In order to generate the singularities of Fig. \ref{fig:singularities}, we start from the first kernel which in the one-loop case reads 
\begin{equation}\label{eq:Lambda0}
\Lambda_{\psi\psi}^{(\alpha\beta),(\gamma\delta)}(P,K)=-\frac{\delta^2\Phi}{\delta S_{\beta\alpha}(P)\delta S_{\gamma\delta}(K)}=-g^2\delta^{\delta\beta}D(P+K)\delta^{\alpha\gamma}\,.
\end{equation}
We iterate it via a first Bethe-Salpeter equation for the four-point function with four fermionic legs:
\begin{eqnarray}\label{eq:BS1}
\Gamma_{\psi\psi}(P,K) & = & \Lambda_{\psi\psi}(P,K)-\int_Q\, \Lambda_{\psi\psi}(P,Q)M(Q)\Gamma_{\psi\psi}(Q,K)\,,\nonumber\\
& = & \Lambda_{\psi\psi}(P,K)-\int_Q\, \Gamma_{\psi\psi}(P,Q)M(Q)\Lambda_{\psi\psi}(Q,K)\,.
\end{eqnarray}
with $M_{(\alpha\beta),(\gamma\delta)}(Q)=S_{\alpha\gamma}(Q)S_{\delta\beta}(Q)$. The indices in (\ref{eq:Lambda0}) have been defined so as to be able to use a generalized matrix product in (\ref{eq:BS1}) (see Appendix \ref{app:matrix}). This Bethe-Salpeter equation resums all the contributions of the type given in Fig. \ref{fig:Lambda_tilde}.
We now use the kernels $\Lambda_{\psi\varphi}$ and $\Lambda_{\varphi\varphi}$ together with $\Gamma_{\psi\psi}$, to define a new kernel as follows:
\begin{eqnarray}\label{eq:Lambda}
\tilde\Lambda_{\varphi\varphi}(P,K) & = & \Lambda_{\varphi\varphi}(P,K)+\int_Q\,\Lambda_{\varphi\psi}(P,Q)M(Q)\Lambda_{\psi\varphi}(Q,K)\nonumber\\
& - & \int_Q\int_R \,\Lambda_{\varphi\psi}(P,Q)M(Q)\Gamma_{\psi\psi}(Q,R)M(R)\Lambda_{\psi\varphi}(R,K)\,,\nonumber\\
& = & \lambda+4g^4\int_Q\,S^{\rm t}(P+Q)M(Q)S(Q+K)\nonumber\\
& - & 4g^4\int_Q\int_R\,S^{\rm t}(P+Q)M(Q)\Gamma_{\psi\psi}(Q,R)M(R)S(R+K)\,,
\end{eqnarray}
where we used the one-loop expressions $\Lambda_{\varphi\varphi}(P,K)=\lambda$, $\Lambda_{\psi\varphi}(P,K)=-2g^2S(P+K)$ and $\Lambda_{\varphi\psi}=\Lambda_{\psi\varphi}^{\rm t}$. The upperscript {\it t} denotes transposition in the generalized index space (see Appendix A). Note that we do not need to use the trace symbol since this is accounted for by the generalized matrix product.
\begin{figure}[htbp]
\begin{center}
\includegraphics[width=8cm]{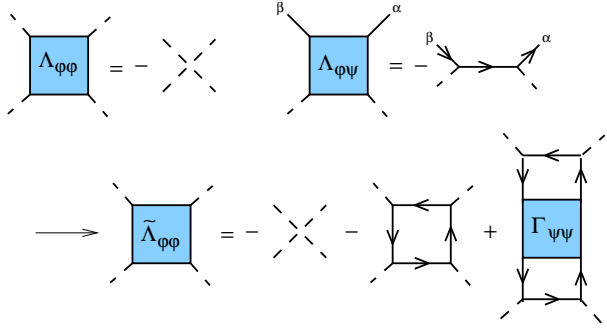}
\caption{New kernel. The factors of 4 present in equation (\ref{eq:Lambda}) and the signs arising from fermion loops are not represented since they are included in the diagrammatic representation.\label{fig:dressed_kernel}}
\end{center}
\end{figure}
This kernel, which we represent in Fig.~\ref{fig:dressed_kernel}, generates all the coupling singularities that we discussed in the previous section via a second Bethe-Salpeter equation for the four-point function with four scalar legs:
\begin{eqnarray}\label{eq:BS}
\Gamma_{\varphi\varphi}(P,K) & = & \tilde\Lambda_{\varphi\varphi}(P,K)-\frac{1}{2}\int_Q\tilde\Lambda_{\varphi\varphi}(P,Q)D^2(Q)\Gamma_{\varphi\varphi}(Q,K)\,,\nonumber\\
& = & \tilde\Lambda_{\varphi\varphi}(P,K)-\frac{1}{2}\int_Q\Gamma_{\varphi\varphi}(P,Q)D^2(Q)\tilde\Lambda_{\varphi\varphi}(Q,K)\,.
\end{eqnarray}
This Bethe-Salpeter equation \footnote{Aside from renormalization in the 2PI framework, this equation plays a role in the context of transport coefficients at high temperature \cite{Aarts:2003bk,Aarts:2004sd,Aarts:2005vc} and also when discussing the so-called Landau-Pomeranchuk-Migdal effect in the context of photon-production in high energy heavy-ion collisions \cite{Serreau:2003wr}.} represented in Fig.~\ref{fig:BS}, contains UV divergences and is renormalized in the very same way than in the case of scalar theories \cite{vanHees:2001ik,Blaizot:2003an}. A detailed analysis is given in section \ref{sec:BSrenorm}.

\section{Renormalization}\label{sec:renormalization}
We now prove that the scalar coupling singularities present in the equations of motion are indeed those encoded in the Bethe-Salpeter equation (\ref{eq:BS}). To this aim we show that, once the equations of motion are expressed in terms of the finite function $\Gamma_{\varphi\varphi}(P,K)$, there are no other divergences. In this section, we illustrate this point at one-loop level. Higher loops are discussed in section \ref{sec:higher_loops}.

\subsection{Renormalized quantities and counterterms}
In order to perform renormalization, it is convenient to express the equations of motion in terms of renormalized (finite) parameters and counterterms (infinite). To make an explicit difference between renormalized and unrenormalized quantities, we label the latter by ``b'' ({\it bare}). We then rescale the fields according to $\psi_{\mbox{\scriptsize b}}= Z^{1/2}_\psi\psi$ and $\varphi_{\mbox{\scriptsize b}}= Z^{1/2}_\varphi\varphi$ so that the action becomes
\begin{equation}
S=\int d^dx\,\left\{Z_\psi\bar\psi(i\slash\hspace{-2.1mm}\partial-m_{\mbox{\scriptsize b}})\psi-g_{\mbox{\scriptsize b}}Z_\psi Z^{1/2}_\varphi\bar{\psi}\psi\varphi+\frac{1}{2}Z_\varphi(\partial\varphi)^2-\frac{1}{2}M_{\mbox{\scriptsize b}}^2Z_\varphi\varphi^2-\frac{1}{4!}\lambda_{\mbox{\scriptsize b}} Z^2_\varphi\varphi^4\right\}.
\end{equation}
Next we introduce renormalized parameters and counterterms by setting
\begin{equation}
Z_\psi m_{\mbox{\scriptsize b}}=m+\delta m\,,\quad Z_\varphi M_{\mbox{\scriptsize b}}^2=M^2+\delta M^2\,,\quad g_{\mbox{\scriptsize b}}Z_\psi Z^{1/2}_\varphi=g+\delta g\,,\quad \lambda_{\mbox{\scriptsize b}} Z_\varphi^2=\lambda+\delta\lambda\,,
\end{equation}
and write the action in the following way:
\begin{eqnarray}
S & = & \int d^dx\,\left\{\bar\psi(i\slash\hspace{-2.1mm}\partial-m)\psi-g\bar{\psi}\psi\varphi+\frac{1}{2}(\partial\varphi)^2-\frac{1}{2}M^2\varphi^2-\frac{1}{4!}\lambda\varphi^4\right.\nonumber\\
& & \hspace{1.5cm}\left.+i\delta Z_\psi \bar\psi\partial\psi-\delta m \bar\psi\psi-\delta g \bar\psi\psi\varphi+\frac{1}{2}\delta Z_\varphi(\partial\varphi)^2-\frac{1}{2}\delta M^2\varphi^2-\frac{1}{4!}\delta\lambda\varphi^4\right\}\,,
\end{eqnarray}
where we have defined $\delta Z_\varphi=Z_\varphi-1$ and $\delta Z_\psi=Z_\psi-1$. We also define the renormalized propagators $S=Z_\psi^{-1}S_{\mbox{\scriptsize b}}$ and $D=Z_\varphi^{-1} D_{\mbox{\scriptsize b}}$ together with the renormalized self-energies (we will indeed show that these are the objects which can be made finite) $\Sigma(P)=S^{-1}(P)+\slashchar{P}-m$ and $\Pi(K)=D^{-1}(K)-K^2-M^2$. We thus have
\begin{equation}
\Sigma(P)=-\slash\hspace{-2.4mm}P\delta Z_\psi+\delta m+Z_\psi\Sigma_{\mbox{\scriptsize b}}(P)\,,\quad\Pi(K)=K^2\delta Z_\varphi+\delta M^2+Z_\varphi\Pi_{\mbox{\scriptsize b}}(K)\,.
\end{equation}
It is then easy to express $Z_\psi\Sigma_{\mbox{\scriptsize b}}$ and $Z_\varphi\Pi_{\mbox{\scriptsize b}}$ in terms of renormalized quantities: In the expressions for $\Sigma_{\mbox{\scriptsize b}}$ and $\Pi_{\mbox{\scriptsize b}}$, one does the replacements $S_{\mbox{\scriptsize b}}\rightarrow S$, $D_{\mbox{\scriptsize b}}\rightarrow D$, $g_{\mbox{\scriptsize b}}\rightarrow g+\delta g$ and $\lambda_{\mbox{\scriptsize b}}\rightarrow \lambda+\delta \lambda$. Thus in the one-loop approximation, the equations of motion read
\begin{eqnarray}\label{eq:motion_ct}
\Sigma(P) & = & -\left(g+\delta g\right)^2\int^{(T)}_Q\,D(Q)S(Q+P)-\slash\hspace{-2.5mm}P\delta Z_{\psi}+\delta m\,,\\
\Pi(K) & = & \frac{1}{2}\left(\lambda+\delta\lambda\right)\int^{(T)}_Q\,D(Q)+\left(g+\delta g\right)^2\int^{(T)}_Q\,S^{\rm t}(Q)S(Q+K)+K^2\delta Z_{\varphi}+\delta M^2\,.\nonumber
\end{eqnarray}
These equations are UV divergent. We shall show that it is possible to absorb consistently all the UV singularities in the counterterms present in (\ref{eq:motion_ct}). In what follows, and to simplify the discussion we study the massless case both at zero and finite temperature. Since we use dimensional regularization $\delta m=0$ and $\delta M^2=0$. Other regularizations, such as using an explicit cut-off, generate linear and quadratic divergences which can be simply absorbed in mass counterterms. As already mentioned, the Yukawa coupling is not renormalized at one-loop level. We then set $\delta g=0$.

Finally the renormalized kernels are obtained by factoring out the appropriate $Z$ factors: $\Lambda_{\psi\psi}=Z_{\psi}^2\Lambda_{\psi\psi}^{\mbox{\scriptsize b}}$, $\Lambda_{\psi\varphi}=Z_{\psi}Z_{\varphi}\Lambda_{\psi\varphi}^{\mbox{\scriptsize b}}$ and $\Lambda_{\varphi\varphi}=Z_{\varphi}^2\Lambda_{\varphi\varphi}^{\mbox{\scriptsize b}}$. They are obtained from the {\it bare} kernels by doing the replacements $S_{\mbox{\scriptsize b}}\rightarrow S$, $D_{\mbox{\scriptsize b}}\rightarrow D$, $g_{\mbox{\scriptsize b}}\rightarrow g+\delta g$ and $\lambda_{\mbox{\scriptsize b}}\rightarrow \lambda+\delta \lambda$. Note that these kernels do not need to be finite since, in the exact theory, they do not correspond to any physical four-point function. It turns out that $\Lambda_{\psi\psi}$ and $\Lambda_{\psi\varphi}$ can be made finite but not $\Lambda_{\varphi\varphi}$ since it contains the counterterm $\delta\lambda$ which is used to renormalize the four-point function $\Gamma_{\varphi\varphi}=Z_{\varphi}^2\Gamma^{\mbox{\scriptsize b}}_{\varphi\varphi}$. At one-loop level, we have
\begin{equation}
\Lambda_{\psi\psi}=-g^2D\,,\quad\Lambda_{\psi\varphi}=-2g^2S\,,\quad\Lambda_{\varphi\varphi}=\lambda+\delta\lambda\,.
\end{equation}
The equations defining $\Gamma_{\psi\psi}=Z_{\psi}^2\Gamma_{\psi\psi}^{\mbox{\scriptsize b}}$, $\tilde\Lambda_{\varphi\varphi}=Z_{\varphi}^2\tilde\Lambda^{\mbox{\scriptsize b}}_{\varphi\varphi}$ and $\Gamma_{\varphi\varphi}=Z_{\varphi}^2\Gamma^{\mbox{\scriptsize b}}_{\varphi\varphi}$ are exactly those given in (\ref{eq:BS1}),  (\ref{eq:Lambda}) and  (\ref{eq:BS}) after one replaces {\it bare} by renormalized quantities.

\subsection{Nested Bethe-Salpeter equations}\label{sec:BSrenorm}
The first step in the renormalization procedure is to renormalize the set of nested Bethe-Salpeter equations. The Bethe-Salpeter equation (\ref{eq:BS1}) defining $\Gamma_{\psi\psi}$ is finite. In the one-loop case, this follows by simple inspection of the diagrams which are resummed by the equation. The only subdiagrams involving loops have four fermionic legs and thus have a negative superficial degree of divergence ($\delta=-2 $). When used to build the kernel $\tilde\Lambda_{\varphi\varphi}$, the function $\Gamma_{\psi\psi}$ generates logarithmic singularities since $\tilde\Lambda_{\varphi\varphi}$ has four scalar legs ($\delta=0$). However the singularities are all overall singularities since there is no divergent subdiagram (all subdiagrams have at least two fermionic legs and thus yield a negative superficial degree of divergence $\delta\leq -1$).

The Bethe-Salpeter equation defining $\Gamma_{\varphi\varphi}$ is the same than the one described in \cite{Blaizot:2003an}. Its renormalization thus follows the very same lines. It amounts to adjusting the scalar coupling counterterm $\delta\lambda$ \footnote{This counterterm does not contain any temperature dependent divergence since the superficial degree of divergence of $\Gamma_{\varphi\varphi}$ is $\delta=0$.} present in $\tilde\Lambda_{\varphi\varphi}(P,K)$ and that originates from the {\it tadpole} diagram. The finite equation for $\Gamma_{\varphi\varphi}(P,K)$ reads:
\begin{eqnarray}
\Gamma_{\varphi\varphi}(P,K)-\Gamma_{\varphi\varphi}(P_*,K_*) & = & \tilde\Lambda_{\varphi\varphi}(P,K)-\tilde\Lambda_{\varphi\varphi}(P_*,K_*)\nonumber\\
& - & \frac{1}{2}\int_Q\Big\{\tilde\Lambda_{\varphi\varphi}(P,Q)-\tilde\Lambda_{\varphi\varphi}(P_*,Q)\Big\}D^2(Q)\Gamma_{\varphi\varphi}(Q,K)\nonumber\\
& - & \frac{1}{2}\int_Q\Gamma_{\varphi\varphi}(P_*,Q)D^2(Q)\Big\{\tilde\Lambda_{\varphi\varphi}(Q,K)-\tilde\Lambda_{\varphi\varphi}(Q,K_*)\Big\}\,,
\end{eqnarray}
where $(P_*,K_*)$ denotes the renormalization point. The finiteness of this equation follows from the fact that $\tilde\Lambda_{\varphi\varphi}(P,Q)-\tilde\Lambda_{\varphi\varphi}(P_*,Q)$ is finite (there is only an overall logarithmic divergence in $\tilde\Lambda_{\varphi\varphi}$) and the property $\tilde\Lambda_{\varphi\varphi}(P,Q)-\tilde\Lambda_{\varphi\varphi}(P_*,Q)\sim 1/Q$ for large $Q$ and fixed $P$ and $P_*$. In pure $\varphi^4$ theory, this property is related to the 2PI structure of the kernel $\tilde\Lambda_{\varphi\varphi}$ \cite{Blaizot:2003an}. Here the kernel $\tilde\Lambda_{\varphi\varphi}$ is not fully 2PI since it arises from a Bethe-Salpeter equation which is two-particle-reducible (2PR) with respect to fermionic lines (see Fig. \ref{fig:Lambda_tilde}). However this does not prevent the asymptotic property of $\tilde\Lambda_{\varphi\varphi}$ to remain true as we discuss in Appendix \ref{app:UV_Lambda}.

\subsection{Asymptotic self-energies}\label{sec:asymptotic}
An helpful quantity in the discussion of UV singularities is the dominant asymptotic piece of each of the self-energies \cite{Blaizot:2003an}. We denote them respectively by $\Sigma_1(P)$ and $\Pi_2(K)$. The labels are used to recall that, for large momenta $\Sigma_1(P)\sim\slash\hspace{-2.5mm}P$ and $\Pi_2(K)\sim K^2$ up to logarithms. These asymptotic pieces satisfy simplified equations of motion obtained after removing all the mass scales (but the renormalization scale $\mu$) in the initial equations of motion (\ref{eq:motion_ct}). For the one-loop case, we obtain
\begin{eqnarray}\label{eq:gap_asymptotic}
\Sigma_1(P) & = & -g^2\int_Q\,D_2(Q)S_1(Q+P)-\slash\hspace{-2.5mm}P\delta Z_{\psi}\,,\nonumber\\
\Pi_2(K) & = & \frac{1}{2}\left(\lambda+\delta\lambda\right)\int_Q\,D_2(Q)+g^2\int_Q\,S_1^{\rm t}(Q)S_1(Q+K)+K^2\delta Z_{\varphi}\,,
\end{eqnarray}
with the asymptotic propagators $S_1(P)=1/(-\slash\hspace{-2.5mm}P+\Sigma_1(P))$ and $D_2(Q)=1/(Q^2+\Pi_2(Q))$. Note that, since we are working in the massless case, the asymptotic equations are nothing but the equations of motion in the vacuum. 

The renormalization of the asymptotic equations of motion is done by means of the field strength counterterms $\delta Z_\psi$ and $\delta Z_\varphi$. The choice
\begin{eqnarray}\label{eq:Z_ct}
\delta Z_\psi & = & -g^2\left.\frac{d}{d\slash\hspace{-2.5mm}P}\int_Q\,D_2(Q)S_1(Q+P)\right|_{P_*}\,,\nonumber\\
\delta Z_\varphi & = & -g^2\left.\frac{d}{dK^2}\int_Q\,S_1^{\rm t}(Q)S_1(Q+K)\right|_{K_*}\,,
\end{eqnarray}
leads to finite equations \footnote{This choice of counterterms is possible since the asymptotic equations of motion are $O(4)$-invariant.}. Note that the field strength counterterms only depend on the dominant asymptotic pieces of the self-energies. Consecuently they are those needed in the initial equations of motion (\ref{eq:motion_ct}). The value of the coupling counterterm $\delta\lambda$ does not play a role since in dimensional regularization $\int_Q D_2(Q)=0$.

Once the dominant pieces are known, we only need to focus on the subleading pieces $\Sigma_{-1}(P)=\Sigma(P)-\Sigma_1(P)$ and $\Pi_0(K)=\Pi(K)-\Pi_2(K)$. The labels are again related to the large momentum behavior: $\Sigma_{-1}(P)\sim 1/\slash\hspace{-2.4mm}P$ and $\Pi_0(K)\sim 1$, up to logarithms. For example, for the fermionic self-energy, the subleading behavior arises by fixing the momentum in one of the lines of the diagram in Fig.~\ref{fig:gap}. The subgraph thus obtained is of degree of divergence $\delta<0$ and according to Weinberg's theorem \cite{Weinb2} decreases at least as $1/\slash\hspace{-2.4mm}P$. In order to obtain equations for $\Sigma_{-1}(P)$ and $\Pi_0(K)$, we use expansions around the asymptotic propagators:
\begin{equation}
S(P)=S_1(P)+\delta S(P)\,,\quad D(K)=D_2(K)+\delta D(K)\,,
\end{equation}
with:
\begin{equation}
\delta S(P)=-S_1(P)\Sigma_{-1}(P)S_1(P)+S_r(P)\,,\quad\delta D(K)=-D_2(K)\Pi_0(K)D_2(K)+D_r(K)\,,
\end{equation}
and $S_r(P)\sim 1/P^5$ and $D_r(K)\sim 1/K^6$ for large momenta. In the massless case, at finite temperature, it is important to keep an infrared regulator which is provided by thermal corrections to the propagator. It is then more convenient to think of an expansion of $S_1$ and $D_2$ around $S$ and $D$ respectively:
\begin{equation}\label{eq:dev}
S_1(P)=S(P)-\delta S(P)\,,\quad D_2(K)=D(K)-\delta D(K)\,,
\end{equation}
with:
\begin{equation}\label{eq:dev2}
\delta S(P)=-S(P)\Sigma_{-1}(P)S(P)+S'_r(P)\,,\quad\delta D(K)=-D(K)\Pi_0(K)D(K)+D'_r(K)\,,
\end{equation}
with $S'_r(P)$ and $D'_r(K)$ keeping the same asymptotic behavior than $S_r(P)$ and $D_r(K)$ respectively.

\subsection{Equations for $\Sigma_{-1}$ and $\Pi_0$}\label{subsec:thermal}
Before writing the equations for $\Sigma_{-1}$ and $\Pi_0$, it is convenient to separate the explicit thermal dependence in the equations of motion. To this aim we perform the Matsubara sums. In the one-loop case, the thermal separation is easily achieved (see Appendix \ref{app:thermal}):
\begin{eqnarray}\label{eq:thermalsep}
\Sigma(P) & = & -g^2\int_Q\,D(Q)S(Q+P)-\slash\hspace{-2.6mm}P\delta Z_{\psi}\nonumber\\
& - & g^2\int_{\tilde Q}\,\sigma_\varphi(\tilde Q)S(\tilde{Q}+P)+g^2\int_{\tilde Q}\,D(P+\tilde Q)\sigma_\psi(\tilde Q)\,,\nonumber\\
\Pi(K) & = & \frac{1}{2}\left(\lambda+\delta\lambda\right)\int_Q\,D(Q)+g^2\int_Q\,S^{\rm t}(Q)S(Q+K)+K^2\delta Z_{\varphi}\nonumber\\
& + & \frac{1}{2}\left(\lambda+\delta\lambda\right)\int_{\tilde Q}\,\sigma_\varphi(\tilde Q)-2g^2\int_{\tilde Q}\,\sigma_\psi^{\rm t}(\tilde Q)S(\tilde Q+K)\,.
\end{eqnarray}
The first lines of each equation are the same functionals of $S$ and $D$ that we would have written in the vacuum. They contain however an implicit thermal dependence via the propagators. The other terms (explicitly depending on temperature) are expressed in terms of $\sigma_\psi(\tilde Q)$ and $\sigma_\varphi(\tilde Q)$ which are defined as follows:
\begin{equation}
\sigma_\psi(\tilde Q)=\epsilon(q_0)n_{|q_0|}\rho_\psi(q_0,q)\,,\quad\sigma_\varphi(\tilde Q)=\epsilon(q_0)N_{|q_0|}\rho_\varphi(q_0,q)\,,
\end{equation}
with $\epsilon(q_0)$ denoting the sign function, $n_{|q_0|}$ and $N_{|q_0|}$ the fermionic and scalar thermal factors (they vanish as $T\rightarrow 0$) and $\rho_\psi(q_0,q)$ and $\rho_\varphi(q_0,q)$ the fermionic and scalar spectral densities respectively (see Appendix \ref{app:thermal}). Finally $\tilde Q$ denotes Minkowskian momentum. Note that if it were not for the counterterm $\delta\lambda$ multiplying the explicitly temperature dependent {\it tadpole} integral, all the explicitly temperature dependent pieces would be finite. For higher loops one would need to exploit the techniques developed in \cite{Blaizot:2004bg} in order to separate explicit thermal dependences.

We now derive equations for the subleading pieces $\Sigma_{-1}$ and $\Pi_0$. We start from the asymptotic equation of motion for the fermion (\ref{eq:gap_asymptotic}) and expand the propagators $S_1$ and $D_2$ by means of (\ref{eq:dev}):
\begin{eqnarray}
\Sigma_1(P) & = & -g^2\int_Q\,D(Q)S(Q+P)-\slash\hspace{-2.4mm}P\delta Z_{\psi}\nonumber\\
& + & g^2\int_Q\,\delta D(Q)S(Q+P)+g^2\int_Q\, D(P+Q)\delta S(Q)-\Sigma_r(P)\,.
\end{eqnarray}
We subtract this equation from (\ref{eq:thermalsep}) and obtain an equation for $\Sigma_{-1}$:
\begin{eqnarray}\label{eq:Sigma}
\Sigma_{-1}(P) & = & -g^2\int_Q\,\delta D(Q)S(Q+P)-g^2\int_Q\, D(P+Q)\delta S(Q)+\Sigma_r(P)\nonumber\\
& - & g^2\int_{\tilde Q}\,\sigma_\varphi(\tilde Q)S(\tilde{Q}+P)+g^2\int_{\tilde Q}\,D(P+\tilde Q)\sigma_\psi(\tilde Q)\,,
\end{eqnarray}
where $\Sigma_r(P)$ is a finite function decreasing as $1/P^3$ as it can be checked on the explicit formula by using power counting and Weinberg's theorem:
\begin{equation}
\Sigma_r(P)=g^2\int_Q\,\delta D(Q)\delta S(Q+P)\,.
\end{equation}

\begin{figure}[htbp]
\begin{center}
\includegraphics[width=10cm]{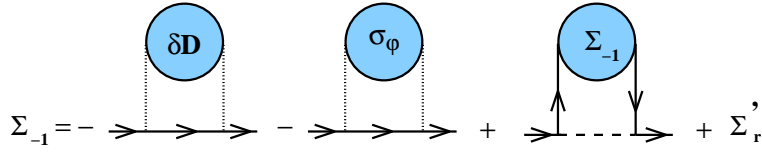}
\caption{Equation of motion for $\Sigma_{-1}$. The dotted lines in the first two diagrams represent the convolution of $\delta D(Q)$ and $S(P+Q)$ or $\sigma_{\varphi}(\tilde Q)$ and $S(P+\tilde Q)$. They do not represent propagators.\label{fig:fermion}}
\end{center}
\end{figure}

\noindent{In order to reveal the self-consistent structure of equation (\ref{eq:Sigma}), we isolate the dominant contribution to $\delta S$ in (\ref{eq:dev2}):}
\begin{eqnarray}\label{eq:Sigma_1}
\Sigma_{-1}(P) & = & -g^2\int_Q\, \delta D(Q)S(Q+P)-g^2\int_{\tilde Q}\,\sigma_\varphi(\tilde Q)S(\tilde{Q}+P)\nonumber\\
& + & g^2\int_Q\, D(P+Q)M(Q)\Sigma_{-1}(Q)+\Sigma'_r(P)\,.
\end{eqnarray}
$\Sigma'_r(P)$ is a finite function, still decreasing as $1/P^3$:
\begin{equation}
\Sigma'_r(P)=\Sigma_r(P)-g^2\int_Q\, D(P+Q)S'_r(Q)+g^2\int_{\tilde Q}\, D(P+{\tilde Q})\,\sigma_\psi(\tilde Q)\,.
\end{equation}  
The last term seems to contradict the $1/P^3$ behavior since the leading contribution at large $P$ is $-g^2D(P)\int_{\tilde Q}\, \,\sigma_\psi(\tilde Q)\sim 1/P^2$. However, as we show in Appendix \ref{app:thermal}, the integral of $\sigma_{\psi}$ vanishes. The asymptotic behavior is thus $\sim 1/P^3$. In Eq. (\ref{eq:Sigma_1}) we have kept the dominant contributions to the asymptotic behavior $1/P$ of $\Sigma_{-1}$. Note that these pieces are finite (this is expected since the equation of motion for the fermionic propagator contains only a field strength singularity which is entirely accounted for by the asymptotic equation of motion). However these dominant pieces, when plugged in the equation of motion for the scalar propagator, generate new singularities which need to be removed. The equation of motion (\ref{eq:Sigma_1}) is represented in Fig. \ref{fig:fermion}.

\begin{figure}[htbp]
\begin{center}
\includegraphics[width=9cm]{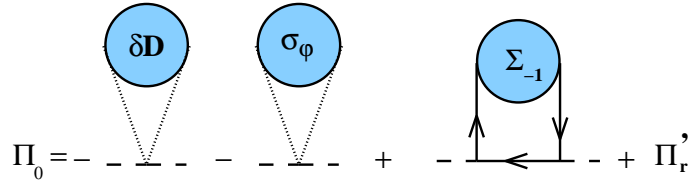}
\caption{Equation of motion for $\Pi_0$. The dotted lines in the first two diagram represent the convolution of $\delta D(Q)$ or $\sigma_{\psi}(\tilde Q)$ with $1$. They do not represent a propagator. Symmetry factors and signs arising from fermion loops are included in the diagrams.\label{fig:scalar}}
\end{center}
\end{figure}

We now proceed in exactly the same way with the equation of motion for the scalar propagator and obtain:
\begin{eqnarray}\label{eq:scalarPi}
\Pi_0(K) & = & \frac{1}{2}\left(\lambda+\delta\lambda\right)\int_Q\,\delta D(Q)+\frac{1}{2}\left(\lambda+\delta\lambda\right)\int_{\tilde Q}\sigma_{\varphi}(\tilde Q)\nonumber\\
& - & 2g^2\int_Q\,S^{\rm t}(K+Q)M(Q)\Sigma_{-1}(Q)+\Pi'_r(K)\,,
\end{eqnarray}
which is represented in Fig. \ref{fig:scalar}. $\Pi'_r(K)$ is given by:
\begin{equation}
\Pi'_r(K)=-g^2\int_Q\,\delta S^{\rm t}(K+Q)\delta S(Q)+2g^2\int_Q\,S^{\rm t}(K+Q)S'_r(Q)-2g^2\int_{\tilde Q}\,\sigma^{\rm t}_{\psi}(\tilde Q)S(K+{\tilde Q})\,.
\end{equation}
Again by power counting and applying Weinberg's theorem we obtain that $\Pi'_r(K)$ is a finite function which goes like $1/K^2$ for large momentum. Unlike the fermionic gap equation, there are logarithmically divergent contributions in equation (\ref{eq:scalarPi}). For the tadpole $\delta=4-4=0$ and for the fermionic insertion $\delta=4-3-1=0$. These logarithmic divergences are those hidden in the Bethe-Salpeter equation for $\Gamma_{\varphi\varphi}$. The game to play is then to try to use the information on the four-point function $\Gamma_{\varphi\varphi}$ to get rid of the singularities.

\subsection{Iterating the equation for $\Sigma_{-1}$}
\begin{figure}[htbp]
\begin{center}
\includegraphics[width=16cm]{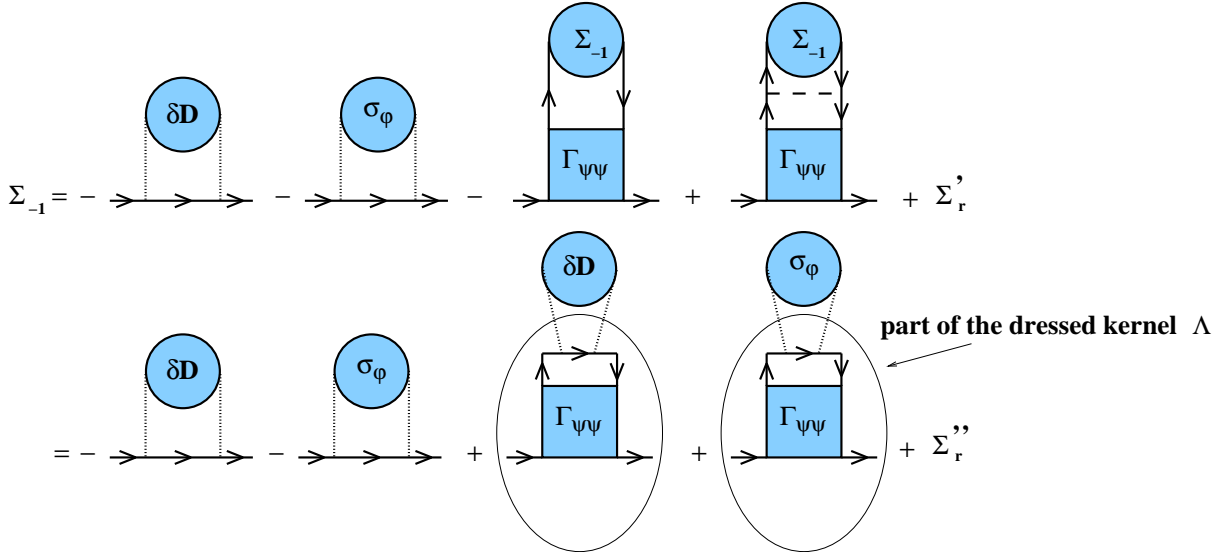}
\caption{Using the first Bethe-Salpeter equation and constructing part of the kernel $\Lambda$.\label{fig:fermion1}}
\end{center}
\end{figure}

We now work on equations (\ref{eq:Sigma_1}) and (\ref{eq:scalarPi}) or Fig.~\ref{fig:fermion} and \ref{fig:scalar}. We start manipulating the equation of motion for the fermionic propagator. We notice that the kernel $\Lambda_{\psi\psi}$ appears in the third term of (\ref{eq:Sigma_1}). We can thus replace it by $\Gamma_{\psi\psi}$ by means of the Bethe-Salpeter equation (\ref{eq:BS1}). This is diagrammatically shown in the first line of Fig.~\ref{fig:fermion1}. We obtain
\begin{eqnarray}\label{eq:toto}
\Sigma_{-1}(P) & = & -g^2\int_Q\, \delta D(Q)S(P+Q)-g^2\int_{\tilde Q}\,\sigma_\varphi(\tilde Q)S(\tilde{Q}+P)+\Sigma'_r(P)\\
& - & \int_Q\, \Gamma_{\psi\psi}(P,Q)M(Q)\Sigma_{-1}(Q)-\int_Q\,\int_K \Gamma_{\psi\psi}(P,K)M(K)\Lambda_{\psi\psi}(K,Q)M(Q)\Sigma_{-1}(Q)\,.\nonumber
\end{eqnarray}
We use the equation of motion (\ref{eq:Sigma_1}) or the Fig.~\ref{fig:fermion} to recognize part of the integrand in (\ref{eq:toto}) (we hide irrelevant terms in $\Sigma''_r$):
\begin{eqnarray}
\Sigma_{-1}(P) & = & -g^2\int_Q\, \delta D(Q)S(P+Q)-g^2\int_{\tilde Q}\,\sigma_\varphi(\tilde Q)S(\tilde{Q}+P)+\Sigma''_r(P)\nonumber\\
& + & g^2\int_Q\int_K\, \Gamma_{\psi\psi}(P,Q)M(Q)S(Q+K)\delta D(K)\nonumber\\
& + & g^2\int_Q\int_{\tilde K}\, \Gamma_{\psi\psi}(P,Q)M(Q)S(Q+{\tilde K})\sigma_\varphi ({\tilde K})\,,
\end{eqnarray}
with:
\begin{equation}\label{eq:Sigma''}
\Sigma''_r(P)=\Sigma'_r(P)-\int_Q\, \Gamma_{\psi\psi}(P,Q)M(Q)\Sigma'_r(Q)\,.
\end{equation}
The properties of $\Sigma''_r(P)$ in the UV are those of $\Sigma'_r(P)$. This manipulation is also represented in Fig.~\ref{fig:fermion1}. Note that we have constructed part of the dressed kernel $\tilde\Lambda_{\varphi\varphi}$ that we need to generate the UV singularities.

\subsection{Iterating the equation for $\Pi_0$}
\begin{figure}[htbp]
\begin{center}
\includegraphics[width=14cm]{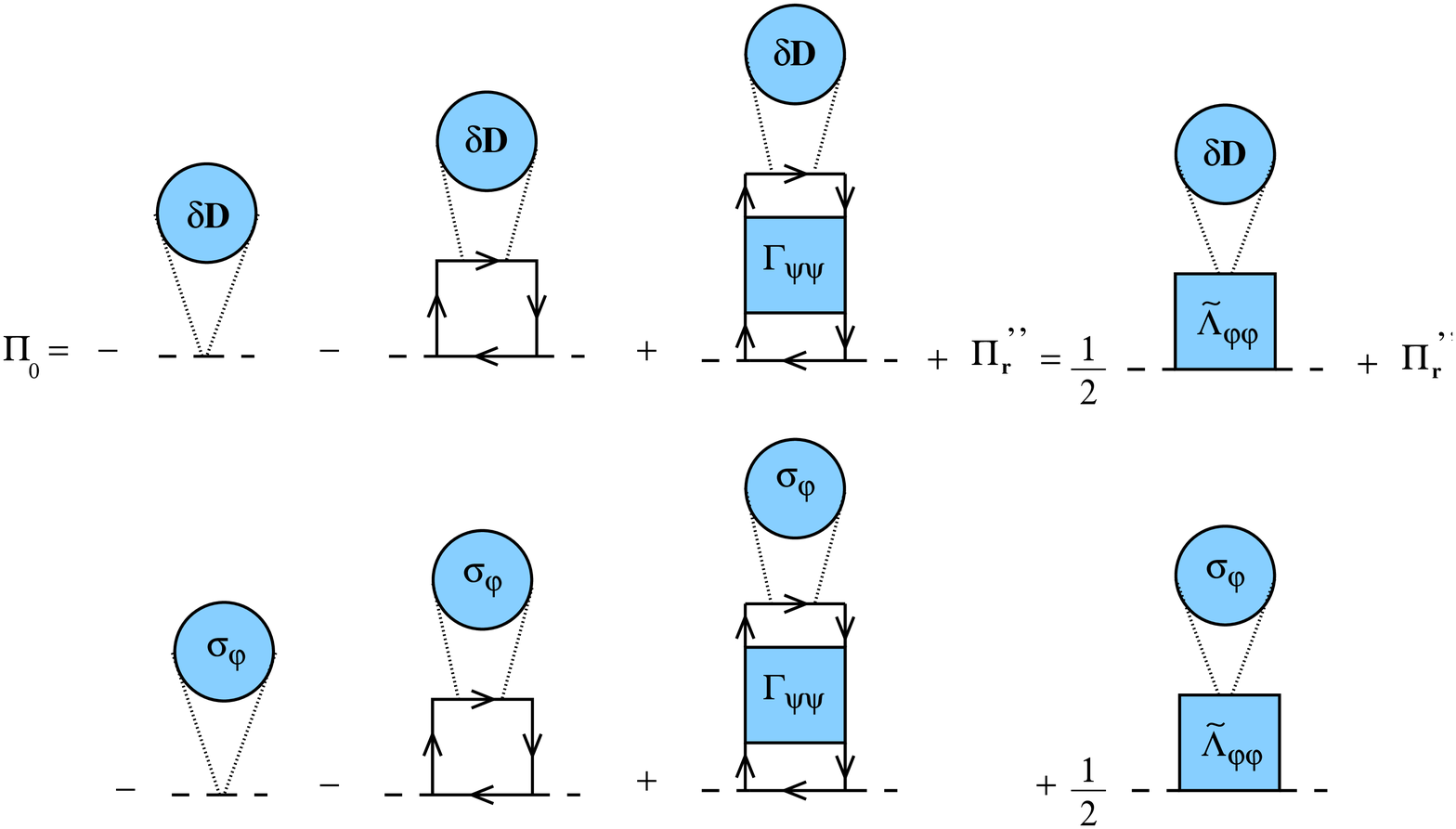}
\caption{Constructing the kernel $\tilde\Lambda_{\varphi\varphi}$.\label{fig:scalar1}}
\end{center}
\end{figure}

Let us now consider the equation for the scalar (\ref{eq:scalarPi}) and plug the expression for $\Sigma_{-1}$that  we have just obtained. Again up to finite quantities which we hide in $\Pi''_r$, we obtain
\begin{eqnarray}\label{eq:Pi0}
\Pi_0(K) & = & \frac{1}{2}\left(\lambda+\delta\lambda\right)\int_Q\,\delta D(Q)+\frac{1}{2}\left(\lambda+\delta\lambda\right)\int_{\tilde Q}\sigma_{\varphi}(\tilde Q)+\Pi''_r(K)\nonumber\\
& + & 2g^4\int_Q\int_P\,S^{\rm t}(K+Q)M(Q)S(Q+P)\delta D(P)\nonumber\\
& + & 2g^4\int_Q\int_{\tilde P}\,S^{\rm t}(K+Q)M(Q)S(Q+{\tilde P})\sigma_{\psi}({\tilde P})\nonumber\\
& - & 2g^4\int_Q\int_K\int_P\,S^{\rm t}(K+Q)M(Q)\Gamma_{\psi\psi}(Q,K)M(K)S(K+P)\delta D(P)\nonumber\\
& - & 2g^4\int_Q\int_K\int_{\tilde P}\,S^{\rm t}(K+Q)M(Q)\Gamma_{\psi\psi}(Q,K)M(K)S(K+\tilde P)\sigma_{\psi}({\tilde P})\,,
\end{eqnarray}
with:
\begin{equation}\label{eq:Pi''}
\Pi''_r(K)=\Pi'_r(K)-2g^2\int_Q\,S^{\rm t}(K+Q)M(Q)\Sigma''_r(Q)\,,
\end{equation}
which again shares the same properties as $\Pi'_r(K)$ in the UV. The manipulation is shown in Fig.~\ref{fig:scalar1}. Equation (\ref{eq:Pi0}) may look complicated but we have indeed reconstructed the kernel $\tilde\Lambda_{\varphi\varphi}$ (see Eq. \ref{eq:Lambda}) and finally
\begin{eqnarray}\label{eq:gap_closed}
\Pi_0(K) & = & \frac{1}{2}\int_Q\tilde\Lambda_{\varphi\varphi}(K,Q)\delta D(Q)+\frac{1}{2}\int_{\tilde Q}\tilde\Lambda_{\varphi\varphi}(K,{\tilde Q})\sigma_\varphi({\tilde Q})+\Pi''_r(K)\,.
\end{eqnarray}

\subsection{Renormalization}
We recover the situation which we had with scalar theories \cite{Blaizot:2003an}. If we now use the Bethe-Salpeter equation (\ref{eq:BS}), we get
\begin{eqnarray}
\Pi_0(K) & = & \frac{1}{2}\int_Q\Gamma_{\varphi\varphi}(K,Q)\delta D(Q)+\frac{1}{2}\int_{\tilde Q}\Gamma_{\varphi\varphi}(K,{\tilde Q})\sigma_{\varphi}(\tilde Q)+\Pi''_r(K)\nonumber\\
& + & \frac{1}{4}\int_Q\int_P\Gamma_{\varphi\varphi}(K,P)D^2(P)\tilde\Lambda_{\varphi\varphi}(P,Q)\delta D(Q)\nonumber\\
& + & \frac{1}{4}\int_{\tilde Q}\int_P\Gamma_{\varphi\varphi}(K,P)D^2(P)\tilde\Lambda_{\varphi\varphi}(P,\tilde Q)\sigma_{\varphi}(\tilde Q)\,,
\end{eqnarray}
The equation of motion (\ref{eq:gap_closed}) allows then to write
\begin{eqnarray}
\Pi_0(K) & = & \frac{1}{2}\int_Q\Gamma_{\varphi\varphi}(K,Q)\delta D(Q)+\frac{1}{2}\int_{\tilde Q}\Gamma_{\varphi\varphi}(K,\tilde Q)\sigma_{\varphi}(\tilde Q)+\Pi''_r(K)\nonumber\\
& + & \frac{1}{2}\int_Q\Gamma_{\varphi\varphi}(K,Q)D^2(Q)\Big\{\Pi_0(Q)-\Pi''_r(Q)\Big\}\,.
\end{eqnarray}
Using the explicit expression for $\delta D(Q)$ given in (\ref{eq:dev2}), we check that the potentially divergent pieces compensate leaving the equation equation:
\begin{equation}\label{eq:ren}
\Pi_0(K)=\Pi''_r(K)+\frac{1}{2}\int_{\tilde Q}\Gamma_{\varphi\varphi}(K,\tilde Q)\sigma_{\varphi}(\tilde Q)+\frac{1}{2}\int_P\Gamma_{\varphi\varphi}(K,P)\Big\{D'_r(P)-D^2(P)\Pi''_r(P)\Big\}\,.
\end{equation}
Using the asymptotic behavior of each of the functions apperaring in the integrands, it is easy to check that this equation does not contain UV divergences. This ends the proof of renormalizability.

\section{Higher loops}\label{sec:higher_loops}
We now show that the general proof follows the same lines than that of the one-loop case. The main idea is that the equations of motion generate the nested Bethe-Salpeter equations which we need to account for the scalar coupling singularities in $\Pi_0$. However we first need to describe the Yukawa coupling singularities in the general case.

\subsection{General approach on the 2PI effective action}
\begin{figure}[htbp]
\begin{center}
\includegraphics[width=16cm]{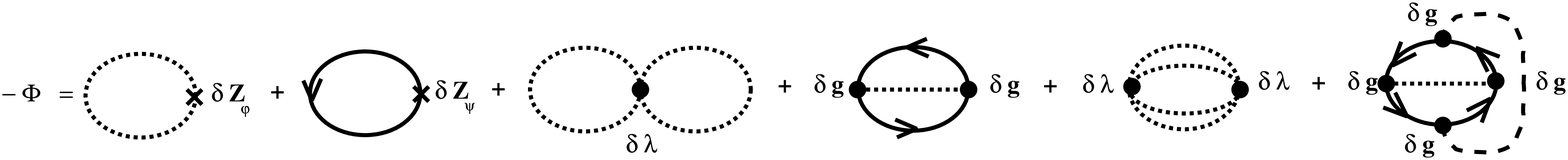}
\caption{$\Phi$ functional expressed in terms of physical parameters and counterterms.\label{fig:Phi2}}
\end{center}
\end{figure}

To understand how the renormalization of the Yukawa coupling works, we first apply a BPH procedure on resummed diagrams. To this aim we make the counterterms explicit in the 2PI effective action. One can easily check that this amounts to replacing {\it bare} propagators by physical propagators, {\it bare} couplings by finite couplings plus counterterms and adding to $\Phi$ two extra diagrams carrying the field strength counterterms. We show $\Phi$ in Fig.~\ref{fig:Phi2}. where we have omitted the {\it renormalized} couplings.
The idea is now to apply a BPH procedure on resummed diagrams. This means drawing boxes which contain divergent diagrams and asking whether there is a structure present in $\Phi$ to account for this singularity. We have three kind of boxes depending on the number of external legs: 2-point boxes are related to field strength singularities, 3-point boxes to the Yukawa coupling renormalization and 4-point boxes to the scalar coupling renormalization.

\begin{figure}[htbp]
\begin{center}
\includegraphics[width=16cm]{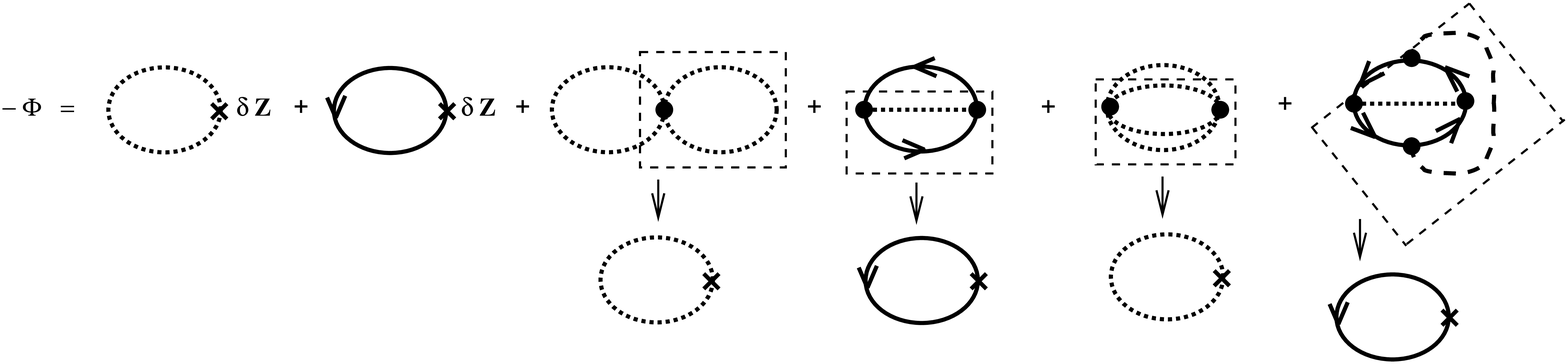}
\caption{BPH procedure to renormalize 2-point singularities. The BPH procedure on resummed graphs is enough to account for this kind of UV singularities.\label{fig:2point}}
\end{center}
\end{figure}

Let us start drawing 2-point boxes as shown in Fig.~\ref{fig:2point}. Because of the 2PI structure, these boxes contain all the lines of a given diagram but one. There are two diagrams waiting to absorb these singularities, namely the first two diagrams in Fig.~\ref{fig:Phi2} carrying the field strength counterterms. This procedure correctly accounts for 2-point singularities and even gives a formula for the field strength counterterms in terms of the propagators  \footnote{One may wonder whether or not the counterterm is temperature dependent: It is not since the divergent piece arises from the asymptotic dominant piece of the propagators.} which generalizes Eq. (\ref{eq:Z_ct}).

\begin{figure}[htbp]
\begin{center}
\includegraphics[width=16cm]{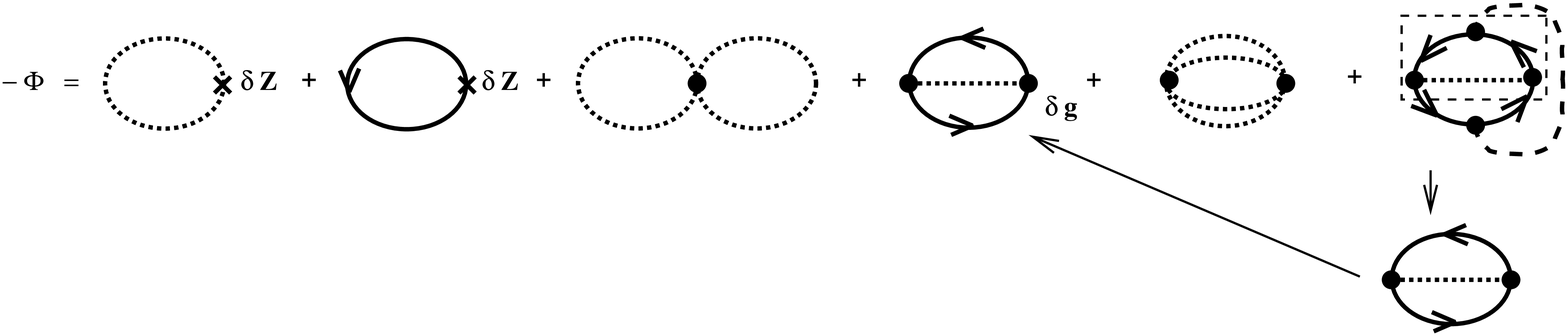}
\caption{BPH procedure to renormalize 3-point singularities. The BPH procedure on resummed graphs is enough to account for this kind of UV singularities.\label{fig:3point}}
\end{center}
\end{figure}

We continue now with 3-point boxes as shown in Fig.~\ref{fig:3point}. We see that we have to go to the last diagram to be able to draw such a 3-point divergent box. Because we are working withing $\Phi$-derivable loop approximations, we know that there is a topology present in $\Phi$ to absorb this singularity, namely the fourth diagram in Fig.~\ref{fig:3point}. This procedure fixes the value of the counterterm $\delta g$ in this diagram. Note that at this order of approximation $\delta g=0$ for the last diagram in Fig.~\ref{fig:Phi2}. This is nothing unusual. When renormalizing at a given order in perturbation theory we use different approximations of the same counterterms in order to construct a finite expression. In particular in the higher loop diagrams $\delta g=0$. From this analysis it is clear that not all $\Phi$-derivable approximations are renormalizable. If one chooses a given diagram to appear in $\Phi$, one has to include all the topologies related to it by the BPH procedure ($\Phi$-derivable loop approximations fulfill this property). Furthermore in order to renormalize such an approximation, one has to allow the counterterms $\delta g$ to be different from one topology to another. This procedure accounts for the divergences that need to be absorbed in the Yukawa coupling renormalization.

\begin{figure}[htbp]
\begin{center}
\includegraphics[width=8cm]{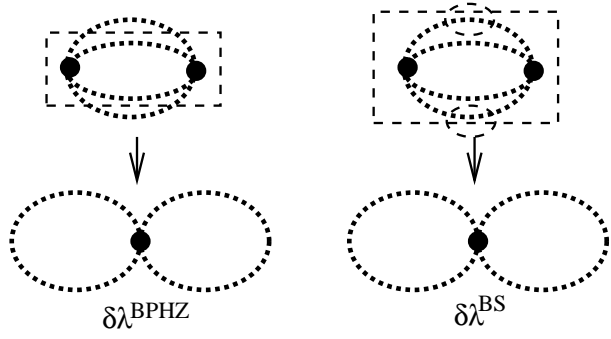}
\caption{The first diagram represents the kind of scalar coupling singularities $\delta\lambda^{\mbox{\scriptsize BPH}}$ accounted for by a BPH procedure on resummed diagrams. This procedure misses singularities. However all these singularities can be accounted for by the Bethe-Salpeter equation and are removed at the level of the effective action by adding a contribution $\delta\lambda^{\mbox{\scriptsize BS}}$ to the total counterterm $\delta\lambda$ in the {\it eight} diagram.\label{fig:BSPhi}}
\end{center}
\end{figure}

Let us finish by mentioning that in the case of 4-scalar singularities, the same remarks made at the level of the equations of motion still hold here. Namely, the BPH approach applied to diagrams with full propagators misses coupling singularities.
\begin{figure}[htbp]
\begin{center}
\includegraphics[width=16cm]{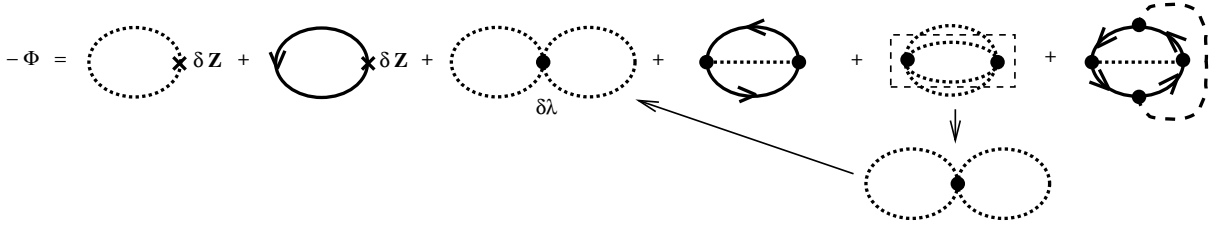}
\caption{BPH procedure to renormalize 4-point singularities. The BPH procedure on resummed graphs is enough to account for this kind of UV singularities. This procedures misses singularities.\label{fig:4point}}
\end{center}
\end{figure}
 However all these extra singularities are, as in the equations of motion, accounted for by a modification of the counterterm in front of the tadpole. This extra contribution accounts for all the coupling singularities generated by the Bethe-Salpeter equation (see Fig. \ref{fig:BSPhi}). To account for all these singularities, we only need to build the correct kernel $\Lambda$. To that aim, in the next section, we look into the equations of motion at any loop order.

\subsection{General nested Bethe-Salpeter equations}

\noindent{The equations of motion can always be written as}
\begin{eqnarray}\label{eq:gap_fermion}
\Sigma_{-1}(P) & = & \frac{1}{2}\int_Q\, \Lambda_{\psi\varphi}(P,Q)\delta D(Q)+\frac{1}{2}\int_{\tilde Q}\, \Lambda_{\psi\varphi}(P,\tilde Q)\sigma_{\varphi}(\tilde Q)+\Sigma'_r(P)\nonumber\\
& - & \int_Q\, \Lambda_{\psi\psi}(P,Q)M(Q)\Sigma_{-1}(Q)\,,
\end{eqnarray}
and
\begin{eqnarray}\label{eq:gap_scalar}
\Pi_0(K) & = & \frac{1}{2}\int_Q\,\Lambda_{\varphi\varphi}(K,Q)\delta D(Q)+\frac{1}{2}\int_{\tilde Q}\,\Lambda_{\varphi\varphi}(K,{\tilde Q})\sigma_{\varphi}(\tilde Q)+\Pi'_r(K)\nonumber\\
& + & \int_Q\,\Lambda_{\varphi\psi}(K,Q)M(Q)\Sigma_{-1}(Q)\,,
\end{eqnarray}
where we have used the basic kernels introduced in section \ref{sec:iterations}. The first step is then to create a new kernel by iterating $\Lambda_{\psi\psi}$ via  a Bethe-Salpeter equation. It is the generalization of the kernel $\Gamma_{\psi\psi}$ that we introduced in the one-loop approximation:
\begin{eqnarray}
\Gamma_{\psi\psi}(P,K) & = & \Lambda_{\psi\psi}(P,K)-\int_Q\Gamma_{\psi\psi}(P,Q)M(Q)\Lambda_{\psi\psi}(Q,K)\,,\nonumber\\
& = & \Lambda_{\psi\psi}(P,K)-\int_Q\Lambda_{\psi\psi}(P,Q)M(Q)\Gamma_{\psi\psi}(Q,K)\,.
\end{eqnarray}
We can then write the equation of motion for the fermion as
\begin{eqnarray}
\Sigma_{-1}(P) & = & \frac{1}{2}\int_Q\, \Lambda_{\psi\varphi}(P,Q)\delta D(Q)+\frac{1}{2}\int_{\tilde Q}\, \Lambda_{\psi\varphi}(P,\tilde Q)\sigma_{\varphi}(\tilde Q)+\Sigma'_r(P)\\
& - & \int_Q\, \Gamma_{\psi\psi}(P,Q)M(Q)\Sigma_{-1}(Q)-\int_Q\int_K\, \Gamma_{\psi\psi}(P,K)M(K)\Lambda_{\psi\psi}(K,Q)M(Q)\Sigma_{-1}(Q)\,.\nonumber
\end{eqnarray}
Using the equation of motion (\ref{eq:gap_fermion}) we obtain
\begin{eqnarray}
\Sigma_{-1}(P) & = & \frac{1}{2}\int_Q\, \Lambda_{\psi\varphi}(P,Q)\delta D(Q)+\frac{1}{2}\int_{\tilde Q}\, \Lambda_{\psi\varphi}(P,\tilde Q)\sigma_{\varphi}(\tilde Q)+\Sigma''_r(P)\nonumber\\
& - & \frac{1}{2}\int_Q\int_K\, \Gamma_{\psi\psi}(P,Q)M(Q)\Lambda_{\psi\varphi}(Q,K)\delta D(K)\nonumber\\
& - & \frac{1}{2}\int_Q\int_{\tilde K}\, \Gamma_{\psi\psi}(P,Q)M(Q)\Lambda_{\psi\varphi}(Q,\tilde K)\delta D(\tilde K)\,.
\end{eqnarray}
We now plug $\Sigma_{-1}$ into the equation of motion for the scalar (\ref{eq:gap_scalar}):
\begin{eqnarray}
\Pi_0(K) & = & \frac{1}{2}\int_Q\,\Lambda_{\varphi\varphi}(K,Q)\delta D(Q)+\frac{1}{2}\int_{\tilde Q}\,\Lambda_{\varphi\varphi}(K,\tilde Q)\sigma_{\varphi}(\tilde Q)+\Pi''_r(K)\nonumber\\
& + & \frac{1}{2}\int_Q\int_P\,\Lambda_{\varphi\psi}(K,Q)M(Q)\Lambda_{\psi\varphi}(Q,P)\delta D(P)\nonumber\\
& + & \frac{1}{2}\int_Q\int_{\tilde P}\,\Lambda_{\varphi\psi}(K,Q)M(Q)\Lambda_{\psi\varphi}(Q,{\tilde P})\sigma_{\varphi}(\tilde P)\nonumber\\
& - & \frac{1}{2}\int_Q\int_K\int_P\,\Lambda_{\varphi\psi}(K,Q)M(Q)\Gamma_{\psi\psi}(Q,K)M(K)\Lambda_{\psi\varphi}(K,P)\delta D(P)\nonumber\\
& - & \frac{1}{2}\int_Q\int_K\int_{\tilde P}\,\Lambda_{\varphi\psi}(K,Q)M(Q)\Gamma_{\psi\psi}(Q,K)M(K)\Lambda_{\psi\varphi}(K,{\tilde P})\sigma_{\varphi}(\tilde P)\,.
\end{eqnarray}
By introducing the kernel
\begin{eqnarray}
\tilde\Lambda_{\varphi\varphi}(K,P) & = & \Lambda_{\varphi\varphi}(K,P)+\int_Q\,\Lambda_{\varphi\psi}(K,Q)M(Q)\Lambda_{\psi\varphi}(Q,P)\nonumber\\
& - & \int_Q\int_R \,\Lambda_{\varphi\psi}(K,Q)M(Q)\Gamma_{\psi\psi}(Q,R)M(R)\Lambda_{\psi\varphi}(R,P)\,,
\end{eqnarray}
the equation of motion for the scalar propagator becomes
\begin{eqnarray}
\Pi_0(K) & = & \frac{1}{2}\int_Q\tilde\Lambda_{\varphi\varphi}(K,Q)\delta D(Q)+\frac{1}{2}\int_{\tilde Q}\tilde\Lambda_{\varphi\varphi}(K,{\tilde Q})({\tilde Q})\sigma_{\varphi}(\tilde Q)+\Pi''_r(K)\,.
\end{eqnarray}
This equation can be renormalized by means of a Bethe-Salpeter equation (\ref{eq:BS}) and leads to the equation (\ref{eq:ren}). $\Pi''_r(K)$ is a finite function, decreasing as $1/K^2$ which ensures the finiteness of equation (\ref{eq:ren}).


\subsection{Renormalization algorithm}
\noindent{The only thing that we need is a rule to construct a finite $\Gamma_{\varphi\varphi}$:}
\begin{itemize}
\item Compute the kernel $\Lambda_{\psi\psi}=-\delta^2\Phi/\delta S^2$.
\item Construct the kernel $\Gamma_{\psi\psi}$ via the first finite Bethe-Salpeter equation:
\begin{eqnarray}\label{eq:bobi}
\Gamma_{\psi\psi}(P,K) & = & \Lambda_{\psi\psi}(P,K)-\int_Q\Gamma_{\psi\psi}(P,Q)M(Q)\Lambda_{\psi\psi}(Q,K)\,,\nonumber\\
& = & \Lambda_{\psi\psi}(P,K)-\int_Q\Lambda_{\psi\psi}(P,Q)M(Q)\Gamma_{\psi\psi}(Q,K)\,.
\end{eqnarray}
\item Compute the kernels $\Lambda_{\psi\varphi}=-2\delta^2\Phi/\delta S\delta D$, $\Lambda_{\varphi\psi}=\Lambda_{\psi\varphi}^{\rm t}$and $\Lambda_{\varphi\varphi}=4\delta^2\Phi/\delta D^2$ and dress the kernel $\Gamma_{\psi\psi}$ into $\tilde\Lambda_{\varphi\varphi}$ by the following construction:
\begin{eqnarray}\label{eq:bobi2}
\tilde\Lambda_{\varphi\varphi}(K,P) & = & \Lambda_{\varphi\varphi}(K,P)\nonumber\\
& + & \int_Q\,\Lambda_{\varphi\psi}(K,Q)M(Q)\Lambda_{\psi\varphi}(Q,P)\nonumber\\
& - & \int_Q\int_R \,\Lambda_{\varphi\psi}(K,Q)M(Q)\Gamma_{\psi\psi}(Q,R)M(R)\Lambda_{\psi\varphi}(R,P)\,.
\end{eqnarray}

\item Construct $\Gamma_{\varphi\varphi}$ via the second finite Bethe-Salpeter equation:
\begin{eqnarray}
\Gamma_{\varphi\varphi}(P,K)-\Gamma_{\varphi\varphi}(P_*,K_*) & = & \tilde\Lambda_{\varphi\varphi}(P,K)-\tilde\Lambda_{\varphi\varphi}(P_*,K_*)\nonumber\\
& - & \frac{1}{2}\int_Q\Big\{\tilde\Lambda_{\varphi\varphi}(P,Q)-\tilde\Lambda_{\varphi\varphi}(P_*,Q)\Big\}D^2(Q)\Gamma_{\varphi\varphi}(Q,K)\nonumber\\
& - & \frac{1}{2}\int_Q\Gamma_{\varphi\varphi}(P_*,Q)D^2(Q)\Big\{\tilde\Lambda_{\varphi\varphi}(Q,K)-\tilde\Lambda_{\varphi\varphi}(Q,K_*)\Big\}\,.
\end{eqnarray}
\end{itemize}
This is equation is finite since the properties of the kernel $\tilde\Lambda_{\varphi\varphi}(P,K)$ revealed in the one-loop case, still hold at higher number of loops as shown in Appendix \ref{app:UV_Lambda}.
 
\begin{figure}[htbp]
\begin{center}
\includegraphics[width=8cm]{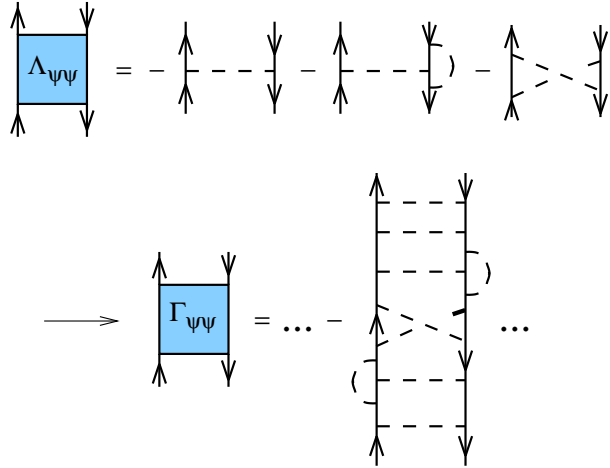}
\caption{Kernel $\Lambda_{\psi\psi}$ and ladder resummation $\Gamma_{\psi\psi}$.\label{fig:lambda0_new}}
\end{center}
\end{figure}

\begin{figure}[htbp]
\begin{center}
\includegraphics[width=14cm]{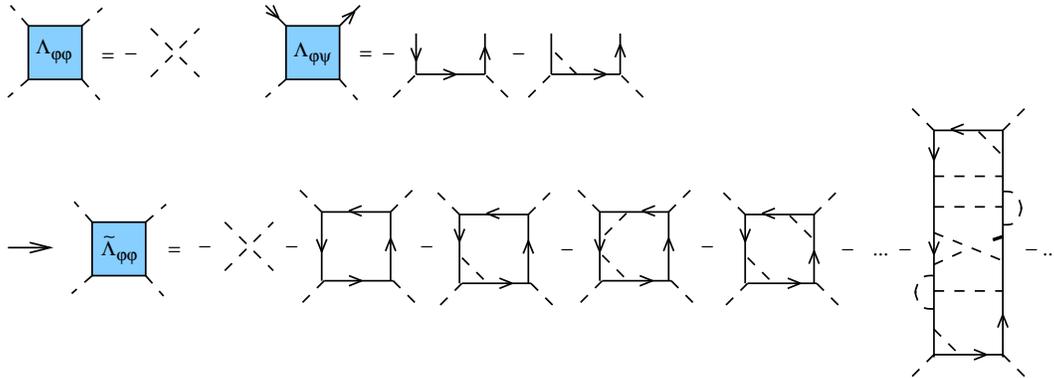}
\caption{Building $\tilde\Lambda_{\varphi\varphi}$ out of $\Gamma_{\psi\psi}$.\label{fig:dressing}}
\end{center}
\end{figure}

As an example we give the divergent structures generated by the inclusion of the fourth diagram in Fig.~\ref{fig:Phi}. The contributions to $\Lambda_{\psi\psi}$ and the corresponding $\Gamma_{\psi\psi}$ are depicted in Fig.~\ref{fig:lambda0_new}.
$\Gamma_{\psi\psi}$ has to be dressed into $\tilde\Lambda_{\varphi\varphi}$ thanks to the diagrams in $\Lambda_{\varphi\varphi}$ and $\Lambda_{\psi\varphi}$. These diagrams are depicted in Fig.~\ref{fig:dressing} together with the corresponding $\tilde\Lambda_{\varphi\varphi}$.
$\Gamma_{\varphi\varphi}$ contains all the ladders build from the rung $\tilde\Lambda_{\varphi\varphi}$.

\section{Conclusion}

We have shown that $\Phi$-derivable approximations are renormalizable in a model where scalar degrees of freedom are coupled to fermions via a Yukawa interaction. If the Yukawa coupling renormalization can be understood by drawing boxes on resummed diagrams, it is not so for the scalar coupling renormalization which amounts to renormalizing a set of nested Bethe-Salpeter equations for the scalar and fermionic four-point functions. We give a systematic rule to construct and renormalize such nested equations at any loop order in the $\Phi$-derivable loop expansion.

This analysis prepares the ground for QED since the power counting rules are strictly the same. Thus one is able to analyse all the UV singularities which may appear in $\Phi$-derivable approximations for Abelian gauge theories. In particular, there are singularities involving four photons. In perturbation theory, the latter are constrained to vanish by the Ward-Takahashi identity. However in any finite loop approximation of the 2PI effective action, these singularities are present since the different channels necessary for the cancelation to occur are not resummed at the same time. This is for example clear with the one-loop topology of Fig. \ref{fig:dressed_kernel} which arises either from the second two-loop diagram in Fig. \ref{fig:Phi} (after cutting one photon line, iterating the equations of motion once and cutting a second photon line, see Fig. \ref{fig:singularity1}) or from the fourth three-loop diagram in the same figure (by directly cutting two photon lines). The way to remove consistently all these spurious singularities will be presented in a subsequent work \cite{BBRS}.

\section*{Acknowledgments}
This work has been supported by the Austrian Science Foundation FWF, project no. P16387-N08. I would like to thank Julien Serreau for fruitful discussions at the early stages of this work and Jean-Paul Blaizot for encouragements and helpful suggestions.

\appendix

\section{Generalized spin indices}\label{app:matrix}
Let us consider a specific example, namely that of the Bethe-Salpeter equation (\ref{eq:BS1}). Writing all the indices, it reads
\begin{eqnarray}
\Gamma_{\psi\psi}^{(\alpha\beta),(\gamma\delta)}(P,K) & = & \Lambda_{\psi\psi}^{(\alpha\beta),(\gamma\delta)}(P,K)-\int_Q\, \Lambda^{(\alpha\beta),(\alpha'\beta')}_{\psi\psi}(P,Q)S_{\alpha'\gamma'}(Q)\Gamma_{\psi\psi}^{(\gamma'\delta'),(\gamma\delta)}(Q,K)S_{\delta'\beta'}(Q).\nonumber\\
\end{eqnarray}
This ordering corresponds to the standard way of writing the diagram by following the fermion lines. We call it {\it cycle} since it first goes up in the diagram and then goes down, following the fermion orientation backwards. There is a more appropriate way to organize indices in the case of Bethe-Salpeter equations. We call this ordering {\it ladder}. It reads
\begin{eqnarray}
\Gamma_{\psi\psi}^{(\alpha\beta),(\gamma\delta)}(P,K) & = & \Lambda_{\psi\psi}^{(\alpha\beta),(\gamma\delta)}(P,K)-\int_Q\, \Lambda^{(\alpha\beta),(\alpha'\beta')}_{\psi\psi}(P,Q)S_{\alpha'\gamma'}(Q)S_{\delta'\beta'}(Q)\Gamma_{\psi\psi}^{(\gamma'\delta'),(\gamma\delta)}(Q,K).\nonumber\\
\end{eqnarray}
Now if we define the four index object $M_{(\alpha'\beta'),(\gamma'\delta')}(Q)=S_{\alpha'\gamma'}(Q)S_{\delta'\beta'}(Q)$ and if we interpret the pair of indices between brackets as generalized indices, we can rewrite the previous equation in terms of a matrix product in this generalized index space. We thus have
\begin{eqnarray}
\Gamma_{\psi\psi}(P,K) & = & \Lambda_{\psi\psi}(P,K)-\int_Q\, \Lambda_{\psi\psi}(P,Q)M(Q)\Gamma_{\psi\psi}(Q,K).
\end{eqnarray}
This matrix notation is perfectly suited for the {\it ladder} structure of the Bethe-Salpeter equation. For instance, iterations of the latter amount to matrix multiplication on the right. This idea can be extended to other situations. For example, if we consider the one fermion loop contributing to the scalar four-point function (see Fig. \ref{fig:Lambda_tilde}), we have to consider the trace 
\begin{eqnarray}
I & = & \int_Q\,S_{\alpha\gamma}(Q)\Lambda_{\psi\varphi}^{(\gamma\delta)}(Q,P) S_{\delta\beta}(Q)\Lambda_{\psi\varphi}^{(\beta\alpha)}(Q,K)\nonumber\\
& = & \int_Q\,\Lambda_{\psi\varphi}^{(\gamma\delta)}(Q,P)M_{(\delta\gamma),(\beta\alpha)}(Q)\Lambda_{\psi\varphi}^{(\beta\alpha)}(Q,K)\,.
\end{eqnarray}
By interpreting $\Lambda^{(\gamma\delta)}_{\psi\varphi}(Q,P)$ as a vector in the generalized index space and by defining the transposed vector by $(\Lambda^{\rm t}_{\psi\varphi})^{(\delta\gamma)}(P,Q)=\Lambda^{(\gamma\delta)}_{\psi\varphi}(Q,P)$, we can write $I$ in a compact form as
\begin{eqnarray}
I & = & \int_Q\,\Lambda^{\rm t}_{\psi\varphi}(P,Q)M(Q)\Lambda_{\psi\varphi}(Q,K)\,,
\end{eqnarray}
without using the trace symbol ``tr''. By definition of the kernels (see Eq. (\ref{eq:kernels})), we have $\Lambda^{\rm t}_{\psi\varphi}=\Lambda_{\varphi\psi}$ so that
\begin{eqnarray}
I & = & \int_Q\,\Lambda_{\varphi\psi}(P,Q)M(Q)\Lambda_{\psi\varphi}(Q,K)\,,
\end{eqnarray}

\section{UV properties of the kernel $\Lambda$}\label{app:UV_Lambda}
We describe here the general properties of the kernel $\tilde\Lambda_{\varphi\varphi}$ at any loop order. $\tilde\Lambda_{\varphi\varphi}$ is constructed from $\Lambda_{\psi\psi}$ in two steps. First one iterates $\Lambda_{\psi\psi}$ through the Bethe-Salpeter equation (\ref{eq:bobi}) and generates $\Gamma_{\psi\psi}$. Next one combines $\Gamma_{\psi\psi}$ together with $\Lambda_{\varphi\varphi}$, $\Lambda_{\varphi\psi}$ and $\Lambda_{\psi\varphi}$ according to Eq. (\ref{eq:bobi2}).

If the approximation belongs to the class of those described in section \ref{sec:higher_loops}, the BPH technique allows to renormalize $\Lambda_{\psi\psi}$. Equation (\ref{eq:bobi}) is then automatically finite since, in the Bethe-Salpeter {\it ladders} that it generates, one cannot find subdiagrams with a non-negative superficial degree of divergence $\delta$. It follows that $\Gamma_{\psi\psi}$ is finite. This kernel $\Gamma_{\psi\psi}$, when combined with $\Lambda_{\varphi\psi}$ and $\Lambda_{\psi\varphi}$ in order to build $\tilde\Lambda_{\varphi\varphi}$, generates divergent structures, since the overall degree of divergence is $\delta=0$. However there are again no divergent subdiagrams. $\tilde\Lambda_{\varphi\varphi}$ contains only an overall divergence and any combination such as $\tilde\Lambda_{\varphi\varphi}(P,Q)-\tilde\Lambda_{\varphi\varphi}(K,Q)$ is finite.

Finally, at fixed $P$ and $K$, the difference $\tilde\Lambda_{\varphi\varphi}(P,Q)-\tilde\Lambda_{\varphi\varphi}(K,Q)$ decreases like $1/Q$ at large $Q$, that is faster than each of the terms of the difference, which go like $\ln Q$. The reason for this is that the leading logarithmic behavior of $\tilde\Lambda_{\varphi\varphi}(P,Q)$ at large $Q$ and fixed $P$ is independent of $P$ and thus vanishes in the difference. To understand this result one applies Weinberg's theorem \cite{Weinb2}. The leading logarithmic behaviour at large $Q$ and fixed $P$ is related to the subgraphs that one can draw on $\tilde\Lambda_{\varphi\varphi}$ which contain the external legs attached to $Q$ and whose degree of divergence is $\delta=0$. Since the two legs attached to $Q$ are scalar lines and since $\delta=0$, the subgraph must have two other scalar external legs. The problem consists then in enumerating all the possible subgraphs with four scalar legs (two of them being attached to $Q$) that one can draw on $\tilde\Lambda_{\varphi\varphi}$. Since $\tilde\Lambda_{\varphi\varphi}$ is 2PI with respect to scalar lines, the only possible subgraph is the diagram itself. The leading logarithmic behavior of this graph at large $Q$ and fixed $P$ is obtained by considering the regime where the momenta of the lines of the graph go to infinity simultaneously. This washes out the presence of $P$ in any of the lines of the diagram and thus the leading asymptotic behavior is independent of $P$.

\section{Thermal separation}\label{app:thermal}
We explain here how to separate the explicit thermal dependence in the case of the one-loop example considered in section \ref{sec:renormalization}.

\subsection{Fermionic self-energy}
We consider first the Matsubara sum appearing in the equation of motion for the fermionic propagator:
\begin{equation}
\int^{(T)}_Q\,D(Q)S(Q+P)=\frac{1}{\beta}\sum_{n}\int_{\bf q}D(Q)S(Q+P)\,,
\end{equation}
with $\int_{\bf q}=\int\frac{d^{d-1} q}{(2\pi)^{d-1}}$. In our notations $Q=(i\omega_n,{\bf q})$ and $P=(i\omega,{\bf p})$, $i\omega_n$ and $i\omega$ being scalar and fermionic Matsubara frequencies respectively. We now introduce the spectral representation for each of the propagators:
\begin{equation}
D(Q)=\int_{q_0}\frac{\rho_{\varphi}(q_0,{\bf q})}{q_0-i\omega_n}\,,\quad S(Q+P)=\int_{r_0}\frac{\rho_{\psi}(r_0,{\bf p}+{\bf q})}{r_0-i\omega-i\omega_n}\,,\nonumber\\
\end{equation}
with $\int_{q_0}=\int_{-\infty}^{+\infty}\frac{dq_0}{2\pi}$. Thus
\begin{equation}\label{eq:Mat}
\int^{(T)}_Q\,D(Q)S(Q+P)=\int_{\bf q}\int_{q_0}\int_{r_0}\rho_{\varphi}(q_0,{\bf q})\rho_{\psi}(r_0,{\bf p}+{\bf q})\frac{1}{\beta}\sum_{n}\frac{1}{(q_0-i\omega_n)(r_0-i\omega-i\omega_n)}\,.
\end{equation}
We now perform the Matsubara sum:
\begin{equation}\label{eq:Mat2}
\frac{1}{\beta}\sum_{n}\frac{1}{(q_0-i\omega_n)(r_0-i\omega-i\omega_n)}=\frac{N_{q_0}-N_{i\omega-r_0}}{r_0-q_0-i\omega}=\frac{N_{q_0}+n_{-r_0}}{r_0-q_0-i\omega}\,.
\end{equation}
The explicit thermal dependence is extracted by means of
\begin{equation}\label{eq:thermal_sep}
N_{q_0}=-\theta(-q_0)+\epsilon(q_0)N_{|q_0|}\,,\quad n_{-r_0}=\theta(r_0)-\epsilon(r_0)n_{|r_0|}\,.
\end{equation}
Plugging this decomposition into Eqs. (\ref{eq:Mat2}) and (\ref{eq:Mat}), the terms with $\theta$ functions can be rewritten as
\begin{equation}
\int_Q\,D(Q)S(Q+P)\,,
\end{equation}
and the terms explicitly dependent on temperature as
\begin{equation}
\int_{\bf q}\int_{r_0}\epsilon(r_0)\rho_{\psi}(r_0,{\bf p}+{\bf q})n_{|r_0|}D(r_0-i\omega,{\bf q})+\int_{\bf q}\int_{q_0}\epsilon(q_0)\rho_{\varphi}(q_0,{\bf q})N_{|q_0|}S(i\omega+q_0,{\bf p}+{\bf q})\,.
\end{equation}
After changing variables $(r_0,{\bf p}+{\bf q})\rightarrow (-q_0,-{\bf q})$ and using $\rho_{\psi}(-q_0,-{\bf q})=\rho_{\psi}(q_0,{\bf q})$ and $D(-q_0,-{\bf q})=D(q_0,{\bf q})$, we obtain
\begin{equation}
-\int_{\bf q}\int_{q_0}\epsilon(q_0)\rho_{\psi}(q_0,{\bf q})n_{|q_0|}D(i\omega+q_0,{\bf p}+{\bf q})+\int_{\bf q}\int_{q_0}\epsilon(q_0)\rho_{\varphi}(q_0,{\bf q})N_{|q_0|}S(i\omega+q_0,{\bf p}+{\bf q})\,.
\end{equation}
Introducing the notations
\begin{equation}
\sigma_{\psi}(q_0,{\bf q})=\epsilon(q_0)n_{|q_0|}\rho_{\psi}(q_0,{\bf q})\,,\quad\sigma_{\varphi}(q_0,{\bf q})=\epsilon(q_0)N_{|q_0|}\rho_{\varphi}(q_0,{\bf q})\,,
\end{equation}
and designating by $\tilde Q=(q_0,{\bf q})$ Minkowskian momentum, we can finally write
\begin{eqnarray}
\int^{(T)}_Q\,D(Q)S(Q+P) & = &\int_Q\,D(Q)S(Q+P)\nonumber\\
& + & \int_{\tilde Q}S(P+\tilde Q)\sigma_{\varphi}(\tilde Q)-\int_{\tilde Q}\sigma_{\psi}(\tilde Q)D(\tilde Q+P)\,.
\end{eqnarray}

\subsection{Scalar self-energy}
The Matsubara sum involved in the scalar self-energy reads
\begin{equation}
\int^{\rm (T)}_Q\,S^{\rm t}(Q)S(Q+K)=\frac{1}{\beta}\sum_{n}\int_{\bf q}\,S^{\rm t}(Q)S(Q+K).
\end{equation}
After using the spectral representation, we obtain:
\begin{equation}
\int^{\rm (T)}_Q\,S^{\rm t}(Q)S(Q+K)=\int_{r_0}\int_{q_0}\int_{\bf q}\,\rho_\psi^{\rm t}(q_0,{\bf q})\rho_\psi(r_0,{\bf q}+{\bf k})\,\frac{1}{\beta}\sum_{n}\frac{1}{(q_0-i\omega_n)(r_0-i\omega-i\omega_n)}\,.
\end{equation}
The Matsubara frequency $i\omega_n$ is now fermionic and we obtain
\begin{equation}
\frac{1}{\beta}\sum_{n}\frac{1}{(q_0-i\omega_n)(r_0-i\omega-i\omega_n)}=\frac{-n_{q_0}+n_{r_0}}{r_0-q_0-i\omega}\,.
\end{equation}
Extracting the thermal dependent part from the thermal factors, we obtain for the thermal contribution
\begin{eqnarray}
-\int_{r_0}\int_{\bf q}\,S^{\rm t}(r_0-i\omega,{\bf q})\sigma_\psi(r_0,{\bf q}+{\bf k})-\int_{q_0}\int_{\bf q}\,\sigma_\psi^{\rm t}(q_0,{\bf q})S(i\omega+q_0,{\bf q}+{\bf k})\,.
\end{eqnarray}
Using a simple change of variables together with the parity properties of $\sigma_\psi$ and $S$, one arrives at
\begin{equation}
\int^{\rm (T)}_Q\,S^{\rm t}(Q)S(Q+K)=\int_Q\,S^{\rm t}(Q)S(Q+K)-2\int_{\tilde Q}\,\sigma_\psi^{\rm t}(\tilde Q)S(\tilde Q+K)\,.
\end{equation}

\subsection{Remark on $\sigma_{\psi}$}
We finish by a remark on $\sigma_{\psi}(q_0,{\bf q})$. It is a matrix that we can decompose into rotation invariant components. In the massless case the decomposition reads
\begin{equation}
\sigma_{\psi}(q_0,{\bf q})=a(q_0,q)\gamma_0+b(q_0,q)\hat{\bf q}^i\gamma^i\,,
\end{equation}
where $a$ and $b$ are respectively odd and even functions in $q_0$. Now if we integrate over $\tilde Q$, we obtain
\begin{equation}
\int_{\tilde Q}\sigma_{\psi}(q_0,{\bf q})=\gamma_0\int_{\tilde Q}a(q_0,q)+\gamma^i\int_{\tilde Q}b(q_0,q)\hat{\bf q}^i=0\,.
\end{equation}

\end{document}